\documentclass[review]{elsarticle}

\usepackage{lineno,hyperref}
\usepackage{amsmath}
\usepackage{amssymb}
\usepackage{algorithmic}
\usepackage{algorithm}
\usepackage{caption}
\usepackage{mathtools}
\usepackage{changes}
\usepackage{multirow}
\usepackage{verbatim}
\usepackage{mathrsfs}
\usepackage{graphicx}
\usepackage{amsmath,amsthm,bm,mathrsfs}

\theoremstyle{plain}

\theoremstyle{definition}

\theoremstyle{remark}
\newtheorem{remark}{Remark}

\modulolinenumbers[5]

\journal{ArXiv.org}

\begin{document}

\begin{frontmatter}

\title{Balanced Truncation via Tangential Interpolation}

\author[uz]{Umair~Zulfiqar}
\author[qs]{Qiu-Yan~Song\corref{mycorrespondingauthor}}
\cortext[mycorrespondingauthor]{Corresponding author}
\ead{qysong@shu.edu.cn}
\author[zx]{Zhi-Hua~Xiao}
\author[vs]{Victor~Sreeram}
\address[uz]{School of Electronic Information and Electrical Engineering, Yangtze University, Jingzhou, Hubei, 434023, China}
\address[qs]{School of Mechatronic Engineering and Automation, Shanghai University, Shanghai, 200444, China}
\address[zx]{School of Statistics and Data Science, Nanjing Audit University, Nanjing, Jiangsu, 211815, China}
\address[vs]{Department of Electrical, Electronic, and Computer Engineering, The University of Western Australia, Perth, 6009, Australia}
\begin{abstract}
This paper examines the construction of $r^{th}$-order truncated balanced realizations via tangential interpolation at $r$ specified interpolation points. It is demonstrated that when the truncated Hankel singular values are negligible—that is, when the discarded states are nearly uncontrollable and unobservable—balanced truncation simplifies to a bi-tangential Hermite interpolation problem at $r$ interpolation points. In such cases, the resulting truncated balanced realization is nearly \(\mathcal{H}_2\)-optimal and thus interpolates the original model at the mirror images of its poles along its residual directions. Additionally, it is shown that existing low-rank balanced truncation algorithms implicitly perform block interpolation to construct a surrogate for the original system, which is subsequently reduced to obtain an approximate truncated balanced realization.

Like standard \(\mathcal{H}_2\)-optimal model reduction, where the interpolation points and tangential directions that yield a local optimum are not known, in balanced truncation as well, the interpolation points and tangential directions required to produce a truncated balanced realization remain unknown. To address this, we propose an iterative tangential interpolation-based algorithm for balanced truncation. This algorithm starts with an initial guess of an \(r^{th}\)-order truncated balanced realization and iteratively refines the interpolation data by performing tangential interpolation at the mirror images of the poles of the current low-rank truncated balanced realization in the residual directions. In each iteration, the rank of the approximated Gramians is incremented by \(r\), followed by low-rank balanced truncation to generate updated interpolation data for the subsequent step. As the Gramian rank increases and the approximation improves, the relative changes in both the interpolation data and approximate Hankel singular values stagnate. Upon convergence, the algorithm yields a low-rank truncated balanced realization that accurately preserves the \(r\) largest Hankel singular values of the original system. An adaptive scheme to automatically select the order \( r \) of the reduced model is also proposed. The algorithm is fully automatic, choosing both the interpolation data and the model order without user intervention. Additionally, an adaptive low-rank solver for Lyapunov equations based on tangential interpolation is proposed, automatically selecting both the interpolation data and the rank without user intervention. The performance of the proposed algorithms is evaluated on benchmark models, confirming their efficacy.
\end{abstract}


\end{frontmatter}

\section{Introduction}
The complexity of modern dynamical systems has been growing rapidly, along with the increasing computational power of the computers used for modeling them. However, simulating, analyzing, and designing these high-order systems poses a significant computational challenge due to limited memory resources. Model Order Reduction (MOR) addresses this issue by producing a reduced-order approximation of the original high-order model. This reduced-order model (ROM) serves as a surrogate for the original system, offering a similar level of accuracy with manageable numerical error. ROMs are more efficient to simulate and analyze, while still retaining the key characteristics of the original system. For further details on this topic, readers are referred to \cite{quarteroni2014reduced,benner2005dimension,obinata2012model,antoulas2005approximation,schilders2008model} and the references therein.

Balanced Truncation (BT) \cite{moore1981principal} is a state-of-the-art classical MOR method, renowned for its accuracy, stability preservation, and well-defined error bounds \cite{enns1984model}. BT retains the most controllable and observable states while truncating the less significant ones. It preserves the dominant Hankel singular values, which represent the contribution of each state to the system's overall energy transfer. Over time, BT has been extended into a broader family of techniques known as Gramian-based MOR methods. For a detailed survey of BT and its extensions, see \cite{gugercin2004survey}. A key limitation of BT is the computational expense of solving Lyapunov equations to compute the Gramians, especially for high-order systems. To mitigate this, low-rank approximations of the Gramians are used, reducing the computational cost significantly \cite{balakrishnan2001efficient}.

Interpolation-based MOR, also known as moment matching, represents another key class of MOR algorithms. In these methods, the ROM matches the original system's transfer function at specific points in the $s$-plane, referred to as interpolation points. These methods are computationally more efficient than BT, as they do not require solving high-order Lyapunov equations. However, the accuracy of the approximation relies heavily on the choice of interpolation points, which is often not straightforward. It has been shown in \cite{gugercin2008h_2,van2008h2} that the $\mathcal{H}_2$-optimal MOR problem, introduced in \cite{wilson1970optimum}, can be viewed as an interpolation problem with specific interpolation points—namely, the mirror images of the ROM’s poles. In the ``Iterative Rational Krylov Algorithm (IRKA)'' \cite{gugercin2008h_2}, the interpolation points are updated iteratively until convergence. IRKA is one of the state-of-the-art MOR algorithms, recognized for its accuracy, automatic selection of interpolation points, and computational efficiency. A more general method for solving the $\mathcal{H}_2$-optimal MOR problem using Sylvester equations was introduced in \cite{xu2011optimal}, and its computational efficiency was later improved in \cite{MPIMD11-11}.   This method, known as the ``Two-sided Iteration Algorithm (TSIA)'', is equivalent to IRKA when both the original system and the ROM have simple poles. For a thorough review of interpolation or moment matching methods, see \cite{antoulas2020interpolatory,astolfi2020model,astolfi2010model} and the references therein.

BT and interpolation algorithms, such as IRKA and TSIA, are generally considered distinct classes of MOR techniques. Recently, efforts have been made to construct the same ROM that BT produces using interpolation methods \cite{ionescu2012balancing,kawano2023gramian}. These attempts are motivated by the fact that interpolation or moment matching methods are computationally more efficient than BT, as they avoid the need to solve high-order Lyapunov equations. While these initial efforts have not led to significant solutions, they have provided a motivation for further research in this direction. This paper addresses the same problem and proposes an iterative tangential interpolation-based algorithm that, like BT, accurately preserves the dominant Hankel singular values.

This paper begins by highlighting the fundamental differences between Gramian-based and interpolation-based MOR approaches. While the projection matrices in both methods satisfy Sylvester equations, these equations exhibit significant structural differences. The problem of retaining the most controllable or observable states while truncating the weakly controllable or observable states is analyzed, revealing its equivalence to a Galerkin projection problem. This reduces to a subset of tangential interpolation conditions associated with $\mathcal{H}_2$-optimal MOR when the truncated states are weakly controllable or observable. Furthermore, we demonstrate that the BT problem simplifies to bi-tangential Hermite interpolation conditions linked to $\mathcal{H}_2$-optimal MOR when the truncated states correspond to negligible Hankel singular values. Additionally, we show that existing low-rank BT algorithms first perform block interpolation to construct a surrogate of the original system, which is then reduced to produce a ROM approximating that obtained through BT.

The specific interpolation points and tangential directions that preserve the singular values of the controllability or observability Gramian are identified only after computing the Gramians. Similarly, the interpolation points and tangential directions that reproduce the BT-equivalent ROM are determined only after performing BT. To address this challenge, we propose two iterative algorithms. The first algorithm begins with an arbitrary initial guess and iteratively refines the interpolation data through successive low-rank approximation steps, progressively increasing the Gramian ranks until convergence while automatically generating the required interpolation data. This process yields low-rank Gramian approximations that preserve their dominant singular values. The second algorithm constructs a ROM that accurately captures the full-order system's dominant Hankel singular values. The effectiveness of both algorithms is demonstrated through benchmark numerical examples, and their computational efficiency is highlighted by the successful reduction of a large-scale model with ten million states.
\section{Preliminaries}
Consider an $n^{th}$-order stable, minimal, linear time-invariant system $H(s)$ with $m$ inputs and $p$ outputs, represented by the state-space realization $(A, B, C)$ as follows:
\begin{align}
H(s)&=C(sI-A)^{-1}B.\label{sys}
\end{align}In this representation, $A \in \mathbb{R}^{n \times n}$ is the state matrix, $B \in \mathbb{R}^{n \times m}$ is the input matrix, and $C \in \mathbb{R}^{p \times n}$ is the output matrix. The state-space equations for the system in (\ref{sys}) are:
\[\dot{x}(t)=Ax(t)+Bu(t),\quad y(t)=Cx(t).\]
The controllability Gramian $P$ and observability Gramian $Q$ for the system in (\ref{sys}) are the solutions of the following Lyapunov equations:
\[AP+PA^T+BB^T=0,\]
\[A^TQ+QA+C^TC=0.\]
Consider an $r^{th}$-order reduced model $H_r(s)$ of $H(s)$, with its state-space realization $(A_r, B_r, C_r)$ described by:
\begin{align}
H_r(s)&=C_r(sI-A_r)^{-1}B_r,\label{sysr}
\end{align}where $A_r \in \mathbb{R}^{r \times r}$, $B_r \in \mathbb{R}^{r \times m}$, and $C_r \in \mathbb{R}^{p \times r}$. This reduced model $H_r(s)$ is obtained through Petrov-Galerkin projection:
\[A_r=W_r^TAV_r,\quad B_r=W_r^TB,\quad C_r=CV_r,\] where $V_r \in \mathbb{R}^{n \times r}$ and $W_r \in \mathbb{R}^{n \times r}$, with the condition $W_r^T V_r = I$. The state-space equations for this reduced model are:
\begin{align}
\dot{x_r}(t)=A_rx_r(t)+B_ru(t),\quad y_r(t)&=C_rx_r.\label{red_sseq}
\end{align}
The controllability Gramian $P_r$ and the observability Gramian $Q_r$ for the reduced model in (\ref{red_sseq}) are the solutions to the following Lyapunov equations:
\[A_rP_r+P_rA_r^T+B_rB_r^T=0,\]
\[A_r^TQ_r+Q_rA_r+C_r^TC_r=0.\]
\begin{remark} If $V_r$ and $W_r$ are replaced by $V_rR$ and $W_rS$, respectively, where $R$ and $S$ are invertible matrices, they will still produce the same ROM $H_r(s)$, but with a different state-space realization \cite{gallivan2004sylvester}.
\end{remark}
\subsection{Review of Interpolation-based MOR \cite{antoulas2020interpolatory,astolfi2010model,gallivan2004sylvester,wolf2014h}}
Let $S_b \in \mathbb{R}^{r \times r}$, $L_b \in \mathbb{R}^{m \times r}$, $S_c \in \mathbb{R}^{r \times r}$, and $L_c \in \mathbb{R}^{p \times r}$ be matrices such that the pairs $(S_b, L_b)$ and $(S_c, L_c)$ are observable. The projection matrices $V_r$ and $W_r$ are then computed by solving the following Sylvester equations:
\begin{align}
AV_r-V_rS_b+BL_b&=0,\label{syl1}\\
A^TW_r-W_rS_c+C^TL_c&=0.\label{syl2}
\end{align}
Next, we decompose $S_b$ and $S_c$ into their eigenvalue decompositions as follows:
\[S_b=T_{sb}^{-1}\begin{bsmallmatrix}\nu_1&\cdots&0\\\vdots&\ddots&\vdots\\0&\cdots&\nu_r\end{bsmallmatrix}T_{sb},\quad S_c=T_{sc}^{-1}\begin{bsmallmatrix}\mu_1&\cdots&0\\\vdots&\ddots&\vdots\\0&\cdots&\mu_r\end{bsmallmatrix}T_{sc}.\]
Here, $\nu_i$ and $\mu_i$ are the right and left interpolation points, respectively, and their associated right and left tangential directions are defined as:
\[\begin{bmatrix}b_1&\cdots&b_r\end{bmatrix}=L_bT_{sb}^{-1},\quad\begin{bmatrix}c_1&\cdots&c_r\end{bmatrix}=L_cT_{sc}^{-1}.\]
The ROM generated by $V_r$ and $W_rS$ (where $S = (V_r^T W_r)^{-1}$) from equations (\ref{syl1}) and (\ref{syl2}) satisfies the following tangential interpolation conditions:
\[H(\nu_i)b_i=H_r(\nu_i)b_i,\quad\textnormal{and}\quad c_i^TH(\mu_i)=c_i^TH_r(\mu_i).\]
It is important to note, as mentioned in Remark 1, that the matrix $S$ does not change the ROM $H_r(s)$ produced by the pair $(V_r, W_r)$, but it ensures the Petrov-Galerkin projection condition $W_r^T V_r = I$ is met. When $\nu_i = \mu_i$, the Hermite interpolation conditions are satisfied as follows:
\[c_i^TH^{\prime}(\nu_i)b_i=c_i^TH_r^{\prime}(\nu_i)b_i.\]
\subsection{Review of $\mathcal{H}_2$-optimal MOR \cite{gugercin2008h_2,van2008h2,xu2011optimal}}
Let $\hat{P}$ and $\hat{Q}$ be the solutions to the following Sylvester equations:
\begin{align}
A\hat{P}+\hat{P}A_r^T+BB_r^T&=0,\label{Ph}\\
A^T\hat{Q}+\hat{Q}A_r+C^TC_r&=0.\label{Qh}
\end{align}
The $\mathcal{H}_2$ norm of the error between $H(s)$ and $H_r(s)$ is given by:
\begin{align}
||H(s)-H_r(s)||_{\mathcal{H}_2}&=\sqrt{CPC^T+2C\hat{P}C_r^T+C_rP_rC_r^T}\nonumber\\
&=\sqrt{B^TQB+2B^T\hat{Q}B_r+B_r^TQ_rB_r}.\nonumber
\end{align}
The following necessary conditions must be met for a local optimum of $||H(s) - H_r(s)||_{\mathcal{H}_2}^2$ \cite{wilson1970optimum}:
\begin{align}
C\hat{P}-C_rP_r&=0,\label{op1}\\
\hat{Q}^TB-Q_rB_r&=0,\label{op2}\\
\hat{Q}^T\hat{P}-Q_rP_r&=0.\label{op3}
\end{align}In \cite{xu2011optimal}, TSIA is introduced, which satisfies these optimality conditions (\ref{op1})–(\ref{op3}) at convergence. Starting with an initial guess for the ROM $(A_r, B_r, C_r)$, TSIA iteratively updates the projection matrices as $V_r = \hat{P}$ and $W_r = \hat{Q}S$, where $S = (\hat{P}^T\hat{Q})^{-1}$, until convergence is achieved.

Assume that both $H(s)$ and $H_r(s)$ have simple poles. In this case, they can be expressed in the following pole-residue form:
\[H(s)=\sum_{i=1}^{n}\frac{l_ir_i^*}{s-\lambda_i},\quad H_r(s)=\sum_{i=1}^{r}\frac{\tilde{l}_{i}\tilde{r}_{i}^*}{s-\tilde{\lambda}_{i}}.\]
The optimality conditions (\ref{op1})–(\ref{op3}) then simplify to the following Hermite interpolation conditions:
\begin{align}
H(-\tilde{\lambda}_i)\tilde{r}_i&=H_r(-\tilde{\lambda}_i)\tilde{r}_i,\label{iop1}\\
\tilde{l}_i^*H(-\tilde{\lambda}_i)&=\tilde{l}_i^*H_r(-\tilde{\lambda}_i)\label{iop2}\\
\tilde{l}_i^*H^{\prime}(-\tilde{\lambda}_i)\tilde{r}_i&=\tilde{l}_i^*H_r^{\prime}(-\tilde{\lambda}_i)\tilde{r}_i.\label{iop3}
\end{align}IRKA \cite{gugercin2008h_2}, a pioneering and efficient $\mathcal{H}_2$-optimal MOR algorithm, ensures that the interpolation conditions (\ref{iop1})–(\ref{iop3}) are satisfied upon convergence. Starting with arbitrary interpolation points and tangential directions, IRKA updates the interpolation data triplet $(\nu_i, b_i, c_i) = (-\tilde{\lambda}_i, \tilde{r}_i, \tilde{l}_i)$ iteratively until convergence is achieved.
\subsection{Review of BT \cite{moore1981principal,enns1984model}}
In BT, the state-space realization $(A,B,C)$ is first transformed into a balanced realization $(A_b,B_b,C_b)$ using a similarity transformation $T_b$, such that: \[A_b=T_b^{-1}AT_b,\quad B_b=T_b^{-1}B,\quad C_b=CT_b.\]
In a balanced realization, the controllability and observability Gramians are equal and diagonal. These Gramians satisfy the following equations:
\[A_b\Sigma+\Sigma A_b^T+B_bB_b^T=0,\]
\[A_b^T\Sigma+\Sigma A_b+C_b^TC_b=0,\]
where the diagonal elements of $\Sigma = \text{diag}(\sigma_1, \dots, \sigma_n)$ are the Hankel singular values of the system. These values are independent of the specific state-space realization and are ordered as $\sigma_i \geq \sigma_{i+1}$. Each $\sigma_i$ represents the square root of the eigenvalue of the product of the controllability and observability Gramians, $PQ$, i.e., $\sigma_i = \sqrt{\lambda_i(PQ)}$.

In BT, the projection matrices are given by $V_r = T_b Z_r^T$ and $W_r = T_b^{-T} Z_r^T$, where $Z_r = \begin{bmatrix} I_r & 0_{r \times (n - r)} \end{bmatrix}$. The truncated balanced realization (TBR) is itself balanced, with equal and diagonal controllability and observability Gramians:
\[A_r\Sigma_r+\Sigma_r A_r^T+B_rB_r^T=0,\]
\[A_r^T\Sigma_r+\Sigma_r A_r+C_r^TC_r=0,\]
where
$\Sigma=\begin{bmatrix}\Sigma_r&0\\0&\Sigma_{n-r}\end{bmatrix}$. The TBR retains the $r$ largest Hankel singular values.

Among the most numerically stable BT algorithms is the balancing square root method \cite{tombs1987truncated}, which proceeds as follows. First, the Gramians \( P \) and \( Q \) are factorized as \( P = L_pL_p^T \) and \( Q = L_qL_q^T \). Next, the singular value decomposition (SVD) is computed:  
\[  
L_q^T L_p = \begin{bmatrix} U_r & U_{n-r} \end{bmatrix} 
\begin{bmatrix} \Sigma_r & 0 \\ 0 & \Sigma_{n-r} \end{bmatrix} 
\begin{bmatrix} R_r^T \\ R_{n-r}^T \end{bmatrix}  
\]  
The reduction matrices \( V_r \) and \( W_r \) are then defined as:  
\[  
V_r = L_pR_r\Sigma_r^{-1/2}, \quad W_r = L_qU_r\Sigma_r^{-1/2}  
\]  
\section{Main Work}
Interpolation-based and Gramians-based MOR share the common feature of being projection techniques. In this section, we begin by highlighting that, while the projection matrices in both approaches satisfy Sylvester equations, there are significant differences in the specific Sylvester equations involved. Next, we explore the conditions under which the projection matrices solve nearly the same Sylvester equations, resulting in the construction of nearly identical ROMs but with different state-space realizations.

Gramians-based MOR first rearranges the system's states based on a certain criterion via a similarity transformation $T$ and then truncates the states considered less significant based on that criterion \cite{antoulas2005approximation,sorensen2002sylvester}. Let the transformed realization be denoted as $(A_t,B_t,C_t)$, obtained through the similarity transformation $T$:
\[
A_t=T^{-1}AT,\quad B_t=T^{-1}B,\quad C_t=CT.\]The state-space equations for the transformed realization $(A_t,B_t,C_t)$ are:
\[
\dot{x}_t(t)=A_tx_t(t)+B_tu(t),\quad y(t)=C_tx_t(t).\]
The controllability Gramian $P_t$ and the observability Gramian $Q_t$ for this realization are related to $P$ and $Q$ as follows:\[P_t=T^{-1}PT^{-T},\quad Q_t=T^TQT.\]
Let us partition $T$ and $T^{-T}$ as:
\[T=\begin{bmatrix}V_r&T_1\end{bmatrix},\quad T^{-T}=\begin{bmatrix}W_r&T_2\end{bmatrix}.\]
Accordingly, the state-space realization $(A_t, B_t, C_t)$ can be partitioned as:
\[A_t=\begin{bmatrix}A_r&A_{12}\\A_{21}&A_{22}\end{bmatrix}, \quad B_t=\begin{bmatrix}B_r\\B_2\end{bmatrix},\quad C_t=\begin{bmatrix}C_r&C_2\end{bmatrix},\]
where $A_{12}=W_r^TAT_1$, $A_{21}=T_2^TAV_r$, $A_{22}=T_2^TAT_1$, $B_2=T_2^TB$, and $C_2=CT_1$. The corresponding partitioned state-space equations are:
\begin{align}
\dot{x}_r(t)&=A_rx_r(t)+A_{12}\tilde{x}(t)+B_ru(t),\nonumber\\
\dot{\tilde{x}}(t)&=A_{21}x_r(t)+A_{22}\tilde{x}(t)+B_2u(t),\nonumber\\
y(t)&=C_rx_r(t)+C_2\tilde{x}(t).\label{partsseq}
\end{align}By neglecting the contribution of $\tilde{x}(t)$ to the system dynamics, the state-space equations (\ref{partsseq}) simplify to the ROM's state-space equations given in (\ref{red_sseq}).

Note that $P_t$ satisfies the following Lyapunov equation:
\begin{align}
A_tP_t+P_tA_t^T+B_tB_t^T&=0.\label{Pt}
\end{align} Substituting $A_t=T^{-1}AT$ and $B_t=T^{-1}B$ into (\ref{Pt}), and then pre-multiplying by $T$ and post-multiplying by $P_t^{-1}$, we observe that $T$ satisfies the following Sylvester equation:
\begin{align}
T^{-1}ATP_t+P_tA_t^T+T^{-1}BB_t^T&=0\nonumber\\
AT+TP_tA_t^TP_t^{-1}+BB_t^TP_t^{-1}&=0\nonumber\\
AT-TS_{b,n}+BL_{b,n}&=0,\nonumber
\end{align}where
\begin{align}
S_{b,n}&=-P_tA_t^TP_t^{-1}=\begin{bmatrix}S_{b,r}&S_{b,12}\\S_{b,21}&S_{b,22}\end{bmatrix}\nonumber\\
L_{b,n}&=B_tP_t^{-1}=\begin{bmatrix}L_{b,r}&L_{b,2}\end{bmatrix}.\nonumber
\end{align}Next, partition $P_t$ and $P_t^{-1}$ as follows:
\begin{align}
P_t&=\begin{bmatrix}P_r&P_{12}\\P_{12}^T&P_{22}\end{bmatrix},&P_t^{-1}&=\begin{bmatrix}P_{i,r}&P_{i,12}\\P_{i,12}^T&P_{i,22}\end{bmatrix}.\nonumber
\end{align} From this, the following relations hold:
\begin{align}
S_{b,r}&=-P_rA_r^TP_{i,r}-P_{12}A_{12}^TP_{i,r}-P_rA_{21}^TP_{i,12}^T-P_{12}A_{22}^TP_{i,12}^T,\nonumber\\
S_{b,12}&=-P_rA_r^TP_{i,12}-P_{12}A_{12}^TP_{i,12}-P_rA_{21}^TP_{i,22}-P_{12}A_{22}^TP_{i,22},\nonumber\\
S_{b,21}&=-P_{12}^TA_r^TP_{i,r}-P_{22}A_{12}^TP_{i,r}-P_{12}^TA_{21}^TP_{i,12}^T-P_{22}A_{22}^TP_{i,12}^T,\nonumber\\
S_{b,22}&=-P_{12}^TA_r^TP_{i,12}-P_{22}A_{12}^TP_{i,12}-P_{12}^TA_{21}^TP_{i,22}-P_{22}A_{22}^TP_{i,22},\nonumber\\
L_{b,r}&=B_r^TP_{i,r}+B_2^TP_{i,12}^T,\nonumber\\
L_{b,2}&=B_2^TP_{i,22}+B_r^TP_{i,12}.\nonumber
\end{align}It is also clear that $V_r$ and $T_1$ satisfy the following Sylvester equations:
\begin{align}
AV_r-V_rS_{b,r}+BL_{b,r}-T_1S_{b,21}&=0,\nonumber\\
AT_1-T_1S_{b,22}+BL_{b,2}-V_rS_{b,12}&=0.\nonumber
\end{align}
Similarly, $Q_t$ is the solution of the following Lyapunov equation:
\begin{align}
A_t^TQ_t+Q_tA_t+C_t^TC_t&=0.\label{Qt}
\end{align}
Substituting $A_t=T^{-1}AT$ and $C_t=CT$ into (\ref{Qt}), and pre-multiplying by $T^{-T}$ and post-multiplying by $Q_t^{-1}$, we find that $T^{-T}$ satisfies the following Sylvester equation:
\begin{align}
A_t^TQ_t+Q_tA_t+C_t^TC_t&=0\nonumber\\
T^TA^TT^{-T}Q_t+Q_tA_t+T^TC^TC_t&=0\nonumber\\
A^TT^{-T}Q_t+T^{-T}Q_tA_t+C^TC_t&=0\nonumber\\
A^TT^{-T}+T^{-T}Q_tA_tQ_t^{-1}+C^TC_tQ_t^{-1}&=0\nonumber\\
A^TT^{-T}-T^{-T}S_{c,b}+C^TL_{c,n}&=0,\nonumber
\end{align}where
\begin{align}
S_{c,n}&=-Q_tA_tQ_t^{-1}=\begin{bmatrix}S_{c,r}&S_{c,12}\\S_{c,21}&S_{c,22}\end{bmatrix}\nonumber\\
L_{c,n}&=C_tQ_t^{-1}=\begin{bmatrix}L_{c,r}&L_{c,2}\end{bmatrix}.\nonumber
\end{align}
Now, partition $Q_t$ and $Q_t^{-1}$ as
\begin{align}
Q_t&=\begin{bmatrix}Q_r&Q_{12}\\Q_{12}^T&Q_{22}\end{bmatrix},&Q_t^{-1}&=\begin{bmatrix}Q_{i,r}&Q_{i,12}\\Q_{i,12}^T&Q_{i,22}\end{bmatrix}.\nonumber
\end{align}From this, the following relations hold:
\begin{align}
S_{c,r}&=-Q_rA_rQ_{i,r}-Q_{12}A_{21}Q_{i,r}-Q_rA_{12}Q_{i,12}^T-Q_{12}A_{22}Q_{i,12}^T,\nonumber\\
S_{c,12}&=-Q_rA_rQ_{i,12}-Q_{12}A_{21}Q_{i,12}-Q_rA_{12}Q_{i,22}-Q_{12}A_{22}Q_{i,22},\nonumber\\
S_{c,21}&=-Q_{12}^TA_rQ_{i,r}-Q_{22}A_{21}Q_{i,r}-Q_{12}^TA_{12}Q_{i,12}^T-Q_{22}A_{22}Q_{i,12}^T,\nonumber\\
S_{c,22}&=-Q_{12}^TA_rQ_{i,12}-Q_{22}A_{21}Q_{i,12}-Q_{12}^TA_{12}Q_{i,22}-Q_{22}A_{22}Q_{i,22},\nonumber\\
L_{c,r}&=C_rQ_{i,r}+C_2Q_{i,12}^T,\nonumber\\
L_{c,2}&=C_2Q_{i,22}+C_rQ_{i,12}.\nonumber
\end{align}
Finally, it is evident that $W_r$ and $T_2$ satisfy the following Sylvester equations:
\begin{align}
A^TW_r-W_rS_{c,r}+C^TL_{c,r}-T_2S_{c,21}&=0,\nonumber\\
A^TT_2-T_2S_{c,22}+C^TL_{c,2}-W_rS_{c,12}&=0.\nonumber
\end{align}
\begin{remark} The similarity transformation $T$ in Gramians-based MOR is constructed such that $P \approx V_r P_r V_r^T$ and $Q \approx W_r Q_r W_r^T$. This works because, in high-order dynamical systems, the Gramians $P$ and $Q$ are typically numerically low-rank. That means most states are nearly uncontrollable and unobservable, with only a few states being significant. By ensuring that $P_r$ and $Q_r$ capture the dominant eigenvalues of $P$ and $Q$, respectively, a good approximation of $H(s)$ is achieved. The Petrov-Galerkin projection condition, $W_r^T V_r = I$, holds in Gramians-based MOR because $T^{-1} T = I$. In contrast, interpolation-based MOR methods do not construct $T$ but instead directly form $V_r$ and $W_r$. The Petrov-Galerkin projection condition, $S^T W_r^T V_r R = I$, is usually enforced by properly choosing the matrices $R$, $S$, or applying the Gram-Schmidt method.
\end{remark}
\subsection{Truncated Controllable Realization (TCR)\label{sub1}}
Let us compute the eigenvalue decomposition of $P$ as $P=T\Lambda_p T^T$, where $\Lambda_p=diag(\lambda_{p,1},\cdots,\lambda_{p,n})$ with $\lambda_{p,i}\geq\lambda_{p,i+1}$. The matrix $T$, when used as a similarity transformation, organizes the states in descending order of the eigenvalues of $P$; for more details, see \cite{ennthesis}. The weakly controllable states (the last $n-r$ states) can then be truncated. Since $T^{-T}=T$, we have $V_r=W_r$, implying that retaining the $r$ most controllable states in the ROM constitutes a Galerkin projection problem. Moreover, in this case, $P_t$ takes the form: $P_t=\begin{bmatrix}\Lambda_{p,r}&0\\0&\Lambda_{p,n-r}\end{bmatrix}$,
where $\Lambda_{p,r}=diag(\lambda_{p,1},\cdots,\lambda_{p,r})$ and $\Lambda_{p,n-r}=diag(\lambda_{p,r+1},\lambda_{p,n})$. Therefore, we have:
\begin{align}
S_{b,r}&=-\Lambda_{p,r}A_r^T\Lambda_{p,r}^{-1},& S_{b,21}&=-\Lambda_{p,n-r}A_{12}^T\Lambda_{p,r}^{-1},\nonumber\\
 L_{b,r}&=B_r^T\Lambda_{p,r}^{-1}.&&\nonumber
\end{align}
Let us now assume that \( P \) is numerically low-rank with a rank of \( r \), meaning that the remaining \( n-r \) states are nearly uncontrollable. This is not a restrictive assumption, as it typically holds for most high-order dynamical systems, where the Gramians exhibit a low-rank property. This property facilitates the computation of \( P \) for large-scale systems and significantly extends the applicability of BT to such systems. For a comprehensive survey of low-rank approximations of Lyapunov equations, refer to \cite{benner2013numerical}.

When the last \( n-r \) states of the realization \( (A_t, B_t, C_t) \) are nearly uncontrollable, i.e., \( \Lambda_{p,n-r} \approx 0 \), the following nearly holds:  
\begin{align}
AV_r+V_r\Lambda_{p,r}A_r^T\Lambda_{p,r}^{-1}+BB_r^T\Lambda_{p,r}^{-1}&\approx0\nonumber\\
AV_r\Lambda_{p,r}+V_r\Lambda_{p,r}A_r^T+BB_r^T&\approx 0.\nonumber
\end{align}Due to the uniqueness of (\ref{Ph}), \( \hat{P} \approx V_r\Lambda_{p,r} \), and thus \( C\hat{P} - C_rP_r \approx 0 \), which corresponds to \( H(-\tilde{\lambda}_i)\tilde{r}_i \approx H_r(-\tilde{\lambda}_i)\tilde{r}_i \). The approximation \( P \approx V_rP_rV_r^T \) is valid in this case since \( P_r \) contains all the significant eigenvalues of \( P \), while the truncated eigenvalues are negligible. However, as the truncated states become more controllable, the approximation \( P \approx V_rP_rV_r^T \) becomes less accurate, and the deviation of the TCR from the interpolation condition (\ref{iop1}) increases, depending on the controllability of the truncated states. For the special case where \( A = A^T \) and \( B = C^T \), the TCR is identical to the TBR.

We have identified the interpolation points and tangential directions needed to capture the \( r \) significant eigenvalues (equivalent to singular values since \( P \) is symmetric) of \( P \). These \( r \) significant eigenvalues/singular values can be preserved by interpolating at the mirror images of the \( r \) poles \( \tilde{\lambda}_i \) of the \( r^{th} \)-order TCR in the direction of its input residuals \( \tilde{r}_i \). Before proceeding further in this discussion, let us consider an example to validate these observations.\\

\begin{figure}[!h]
  \centering
  \includegraphics[width=8cm]{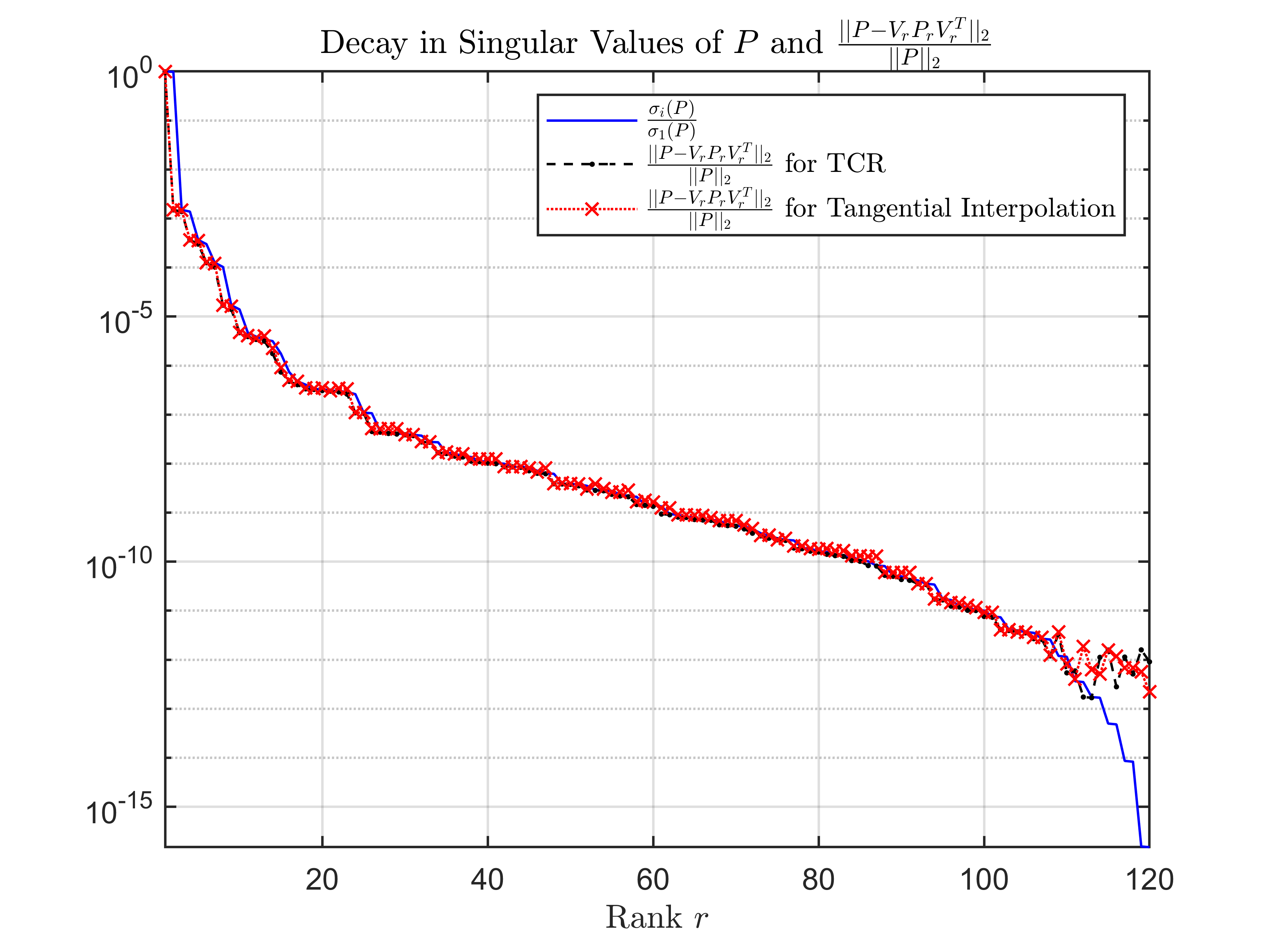}
  \caption{Decay in singular values of $P$ and the relative error $\frac{||P-V_rP_rV_r^T||_2}{||P||_2}$}\label{fig01}
\end{figure}

\textbf{Example 1: CD Player} Consider the \(120^\text{th}\)-order CD player model with $2$ inputs and $2$ outputs, taken from the benchmark collection of models for testing MOR algorithms in \cite{chahlaoui2005benchmark}. This model is chosen because the dominant singular values of \( P \) are exceptionally large in magnitude. Our goal is to examine the effect of violating the assumption that truncated states correspond to small singular values of \( P \). The singular values of \( P \), normalized by the largest (first) singular value, are plotted on a logarithmic scale in Figure \ref{fig01}. The plot shows that the singular values of \( P \) decay rapidly relative to the largest singular value. Subsequently, TCRs of orders \(1\) to \(120\) are constructed, and the relative error \( \frac{\|P - V_rP_rV_r^T\|_2}{\|P\|_2} \) is plotted on the same figure. The relative error decays rapidly, following the trend of the singular values of \( P \). Finally, tangential interpolation is performed using \( -\tilde{\lambda}_i \) (from the TCRs) as interpolation points and \( \tilde{r}_i \) (from the TCRs) as tangential directions. The relative error \( \frac{\|P - V_rP_rV_r^T\|_2}{\|P\|_2} \) for tangential interpolation is also plotted in the same figure, showing nearly identical results to those obtained using TCR. The six largest singular values of \( P \) are as follows:   $1.1715\times10^{6}$, $1.1483\times 10^{6}$, $1.7582\times 10^{3}$, $1.6216\times 10^{3}$, $429.6764$, and $ 351.6459$. Despite the large singular values associated with the truncated states, Figure \ref{fig01} shows that interpolation at the mirror images of the poles of TCRs in their residual directions achieves accuracy nearly identical that of TCRs. The next section will demonstrate through additional benchmark models that this interpolation approach maintains high accuracy even when the truncated singular values of P have large magnitudes.
\subsection{Truncated Observable Realization TOR\label{sub2}}
Let us compute the eigenvalue decomposition of $Q$ as $Q=T\Lambda_q T^T$, where $\Lambda_q=diag(\lambda_{q,1},\cdots,\lambda_{q,n})$ with $\lambda_{q,i}\geq\lambda_{q,i+1}$. The matrix $T$, when used as a similarity transformation, arranges the states in descending order of the eigenvalues of $Q$; for more details, see \cite{ennthesis}. The weakly observable states (the last $n-r$ states) can then be truncated. Since $T^{-T}=T$, we have $V_r=W_r$, implying that retaining the $r$ most observable states in the ROM is a Galerkin projection problem. Furthermore, in this case, $Q_t$ takes the form:

$Q_t=\begin{bmatrix}\Lambda_{q,r}&0\\0&\Lambda_{q,n-r}\end{bmatrix}$, where $\Lambda_{q,r}=diag(\lambda_{q,1},\cdots,\lambda_{q,r})$ and $\Lambda_{q,n-r}=diag(\lambda_{q,r+1},\lambda_{q,n})$. Therefore, we have:
\begin{align}
S_{c,r}&=-\Lambda_{q,r}A_r\Lambda_{q,r}^{-1},& S_{c,21}&=-\Lambda_{q,n-r}A_{21}\Lambda_{q,r}^{-1},\nonumber\\
 L_{c,r}&=C_r\Lambda_{q,r}^{-1}.&&\nonumber
\end{align}
Let us assume that $Q$ is numerically low rank with a numerical rank of $r$, meaning the remaining $n-r$ states are nearly unobservable. When the last \( n-r \) states of the realization \((A_t, B_t, C_t)\) are nearly unobservable (i.e., \(\Lambda_{q,n-r} \approx 0\)), leading to \(S_{c,21} \approx 0\), the following nearly holds:
\begin{align}
A^TW_r+W_r\Lambda_{q,r}A_r\Lambda_{q,r}^{-1}+C^TC_r\Lambda_{q,r}^{-1}&\approx0\nonumber\\
A^TW_r\Lambda_{q,r}+W_r\Lambda_{q,r}A_r+C^TC_r&\approx0.\nonumber
\end{align}Due to the uniqueness of (\ref{Qh}), \(\hat{Q} \approx W_r \Lambda_{q,r}\), and thus \(\hat{Q}^T B - Q_r B_r \approx 0\), which corresponds to \(\tilde{l}_i^* H(-\tilde{\lambda}_i) \approx \tilde{l}_i^* H_r(-\tilde{\lambda}_i)\). The approximation \(Q \approx W_r Q_r W_r^T\) is accurate in this case because \(Q_r\) contains all the significant eigenvalues of \(Q\), while the truncated eigenvalues are negligible. However, as the truncated states become more observable, the approximation \(Q \approx W_r Q_r W_r^T\) loses accuracy, and the deviation of the TOR from the interpolation condition (\ref{iop2}) increases—depending on the observability of the truncated states. For the special case where \(A = A^T\) and \(B = C^T\), the TOR is identical to the TCR and the TBR.  

We have identified the interpolation points and tangential directions necessary to capture the $r$ significant eigenvalues (which are also equal to the singular values since $Q$ is symmetric) of $Q$. These $r$ eigenvalues/singular values of $Q$ can be preserved by interpolating at the mirror images of the $r$ poles $\tilde{\lambda}_i$ of the $r^{th}$-order TOR in the direction of its output residuals $\tilde{l}_i^*$.
\subsection{Truncated Balanced Realization (TBR)\label{sub3}}
For the TBR, \( P_t = Q_t = \begin{bmatrix} \Sigma_r & 0 \\ 0 & \Sigma_{n-r} \end{bmatrix} \), both the controllability and observability Gramians are diagonal, combining the two previously discussed cases. Thus:
\begin{align}
S_{b,r}&=-\Sigma_{r}A_r^T\Sigma_{r}^{-1},& S_{b,21}&=-\Sigma_{n-r}A_{12}^T\Sigma_{r}^{-1},\nonumber\\
 L_{b,r}&=B_r^T\Sigma_{r}^{-1}.&&\nonumber\\
S_{c,r}&=-\Sigma_{r}A_r\Sigma_{r}^{-1},& S_{c,21}&=-\Sigma_{n-r}A_{21}\Sigma_{r}^{-1},\nonumber\\
 L_{c,r}&=C_r\Sigma_{r}^{-1}.&&\nonumber
\end{align}
Let us assume that only \( r \) Hankel singular values are significant. This assumption is reasonable since BT is most effective for systems with rapidly decaying Hankel singular values. For systems like Example 3.2 in \cite{karachalios2023data}, where all Hankel singular values are identical, BT is not an appropriate MOR method.  When the truncated states correspond to insignificant Hankel singular values (\( \Sigma_{n-r} \approx 0 \)), we obtain \( S_{b,21} \approx 0 \) and \( S_{c,21} \approx 0 \). Consequently, \( V_r \) and \( W_r \) nearly satisfy:
\[AV_r+V_r\Sigma_rA_r^T\Sigma_r^{-1}+BB_r\Sigma_r^{-1}\approx 0\]
\[A^TW_r-W_r\Sigma_rA_r\Sigma_r^{-1}+C^TC_r\Sigma_r^{-1}\approx 0.\]
This further simplifies to:
\[AV_r\Sigma_r-V_r\Sigma_rA_r+BB_rT\approx 0\]
\[A^TW_r\Sigma_r-W_r\Sigma_rA_r+C^TC_r\approx 0.\]
Due to the uniqueness of (\ref{Ph}) and (\ref{Qh}), \( \hat{P} \approx V_r \Sigma_r \) and \( \hat{Q} \approx W_r \Sigma_r \). Moreover, since \( W_r^T V_r = I \), the TBR nearly satisfies the optimality conditions (\ref{op1})–(\ref{op3}) and the interpolation conditions (\ref{iop1})–(\ref{iop3}). If the truncated states are associated with significant Hankel singular values, the deviation from these interpolation conditions increases.

We have identified the interpolation points and tangential directions required to capture the $r$ significant Hankel singular values of $H(s)$. These $r$ significant Hankel singular values can be preserved by interpolating at the mirror images of the $r$ poles $\tilde{\lambda}_i$ of the $r^{th}$-order TBR in the directions of its input residuals $\tilde{r}_i$ and output residuals $\tilde{l}_i^*$. Before proceeding further, let us consider an example to validate these observations.\\

\begin{figure}[!h]
  \centering
  \includegraphics[width=8cm]{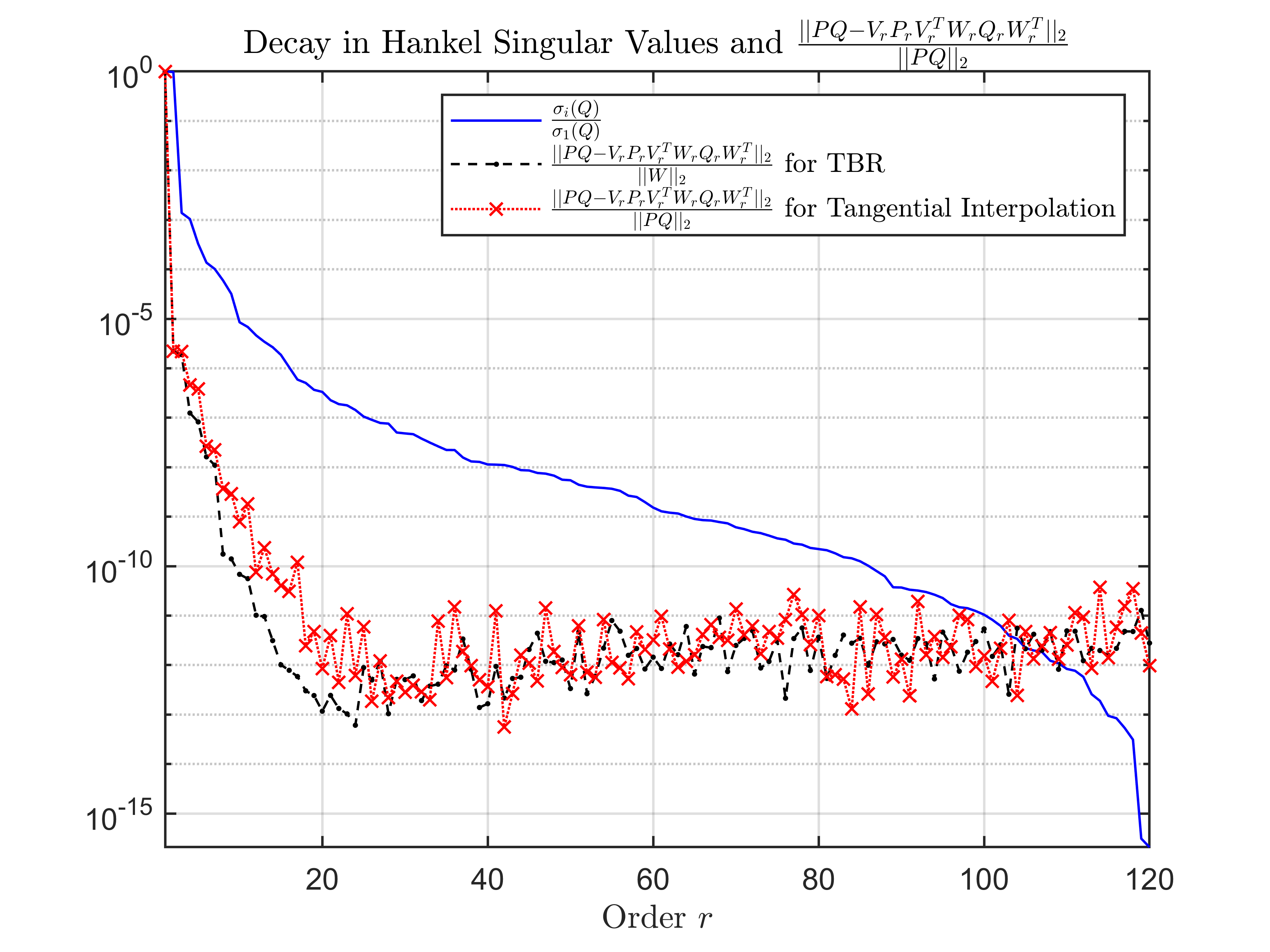}
  \caption{Decay in Hankel singular values of $H(s)$ and the relative error $\frac{||PQ-V_rP_rV_r^TW_rQ_rW_r^T||_2}{||PQ||_2}$}\label{fig03}
\end{figure}

\textbf{Example 1: CD Player (Continued)} The six largest Hankel singular values of this model are as follows: $1.1715\times10^{6}$, $1.1483\times10^{6}$, $1.7386\times 10^{3}$, $1.6016\times10^{3}$, $406.9641$, and $329.3257$. Figure \ref{fig03} shows the Hankel singular values of $H(s)$, normalized by the largest (first) Hankel singular value, plotted on a logarithmic scale. The values decay rapidly relative to the dominant singular value. Subsequently, TBRs of orders $1-120$ are constructed, and the relative error $\frac{||PQ-V_rP_rV_r^TW_rQ_rW_r^T||_2}{||PQ||_2}$ is plotted on the same figure. It is observed that the relative error $\frac{||PQ-V_rP_rV_r^TW_rQ_rW_r^T||_2}{||PQ||_2}$ decays rapidly along with the decay in the Hankel singular values of $H(s)$. Finally, tangential interpolation is performed using $-\tilde{\lambda}_i$ of the TBR as interpolation points and $\tilde{r}_i$ and $\tilde{l}_i^*$ of the TBR as tangential directions. The relative error $\frac{||PQ-V_rP_rV_r^TW_rQ_rW_r^T||_2}{||PQ||_2}$ achieved through tangential interpolation is nearly identical to that obtained using TBR. Again, note that even when truncated states correspond to large Hankel singular values, interpolation at the mirror images of TBR poles in the direction of their associated residuals achieves nearly identical accuracy.
\subsection{Working Principle of Existing Low-rank Methods}\label{sub4}
In the previous subsection, we observed that the TCR, TOR, and TBR reduce to a tangential interpolation problem when the Gramians are low-rank. In this section, we highlight that low-rank methods, such as Krylov-subspace-based methods \cite{jaimoukha1994krylov} and the alternating-direction implicit (ADI) method \cite{benner2013efficient}, perform block interpolation to approximate TCR, TOR realizations, and TBR.

The Krylov-subspace-based methods \cite{jaimoukha1994krylov} provide low-rank approximations of $P$ and $Q$ as follows: $P \approx (V_kZ_p)(V_kZ_p)^T$ and $Q \approx (W_kZ_q)(W_kZ_q)^T$, where $V_k^TV_k = I$ and $W_k^TW_k = I$. If $r$ interpolation points are used for these approximations, the ranks of $Z_p$ and $Z_q$ are $k = rm$ and $k = rp$, respectively. Similarly, the low-rank Cholesky factor-based ADI (LRCF-ADI) method \cite{benner2013efficient} generates the following approximations of $P$ and $Q$: $P \approx V_kV_k^T$ and $Q \approx W_kW_k^T$. If $r$ ADI shifts are used, the ranks of $V_k$ and $W_k$ are $k = rm$ and $k = rp$, respectively. Both the Krylov-subspace-based methods \cite{jaimoukha1994krylov} and the LRCF-ADI method \cite{benner2013efficient} satisfy the following property:
\[\underset {i=1,\cdots,r}{\textnormal{span}}\{(\nu_iI-A)^{-1}B\}\subset \textnormal{Ran}(V_k),\]
\[\underset {i=1,\cdots,r}{\textnormal{span}}\{(\mu_iI-A^T)^{-1}C^T\}\subset \textnormal{Ran}(W_k),\]
where $\nu_i$ and $\mu_i$ are interpolation points in the Krylov-subspace-based method, and $-\nu_i$ and $-\mu_i$ are shifts in the LRCF-ADI method; cf. \cite{wolf2016adi}. 

For low-rank TCR, the balancing square root algorithm can be adapted as follows:
\[Z_p^TV_k^TV_kZ_p=\begin{bmatrix}U_r&U_{k-r}\end{bmatrix}\begin{bmatrix}S_r&0\\0&S_{k-r}\end{bmatrix}\begin{bmatrix}U_r^T\\U_{k-r}^T\end{bmatrix}.\]
In the case of the Krylov-subspace-based approach, $Z_p^TV_k^TV_kZ_p = Z_p^TZ_p$, whereas for the LRCF-ADI method, $Z_p^TV_k^TV_kZ_p = V_k^TV_k$. Setting $V_r = Z_pU_rS_r^{-\frac{1}{2}}$ ensures that $V_r^TV_r = I$. The low-rank TCR is then obtained as:
\begin{align}
A_r=V_r^T\big(V_k^TAV_k\big)V_r,\quad B_r=V_r^T\big(V_k^TB\big),\quad C_r=\big(CV_k\big)V_r.\nonumber
\end{align}It becomes clear that the low-rank TCR can be seen as a two-step procedure. First, a $k^{th}$-order interpolant $CV_k(sI-V_k^TAV_k)^{-1}V_k^TB$ is constructed, which interpolates $H(s)$ at $\nu_i$. Then, it is reduced to order $r$ to obtain the low-rank TCR. When $m = 1$, $k = r$, and in this case, the low-rank TCR becomes an interpolant of $H(s)$ at $\nu_i$.

Similarly, for low-rank TOR, the balancing square root algorithm can be adapted as follows:
\[Z_q^TW_k^TW_kZ_q=\begin{bmatrix}U_r&U_{k-r}\end{bmatrix}\begin{bmatrix}S_r&0\\0&S_{k-r}\end{bmatrix}\begin{bmatrix}U_r^T\\U_{k-r}^T\end{bmatrix}.\]
In the Krylov-subspace-based approach, $Z_q^TW_k^TW_kZ_q = Z_q^TZ_q$, whereas for the LRCF-ADI method, $Z_q^TW_k^TW_kZ_q = W_k^TW_k$. Now, set $W_r = Z_qU_rS_r^{-\frac{1}{2}}$ such that $W_r^TW_r = I$. The low-rank TOR is then obtained as:
\begin{align}
A_r=W_r^T\big(W_k^TAW_k\big)W_r,\quad B_r=W_r^T\big(W_k^TB\big),\quad C_r=\big(CW_k\big)W_r.\nonumber
\end{align}It becomes evident that the low-rank TOR can also be interpreted as a two-step procedure. First, a $k^{th}$-order interpolant $CW_k(sI-W_k^TAW_k)^{-1}W_k^TB$ is constructed, which interpolates $H(s)$ at $\mu_i$. Then, it is reduced to order $r$ to obtain the low-rank TOR. When $p = 1$, $k = r$, and in this case, the low-rank TOR becomes an interpolant of $H(s)$ at $\mu_i$.

Let us assume, for the sake of discussion, that both $V_k$ and $W_k$ have the same column rank. The low-rank TBR can then be obtained using the balancing square root algorithm as follows:
\[Z_q^TW_k^TV_kZ_p=\begin{bmatrix}U_r&U_{k-r}\end{bmatrix}\begin{bmatrix}S_r&0\\0&S_{k-r}\end{bmatrix}\begin{bmatrix}R_r^T\\R_{k-r}^T\end{bmatrix}.\]
Set $V_r = Z_pR_rS_r^{-\frac{1}{2}}$ and $W_r = Z_qU_rS_r^{-\frac{1}{2}}$ such that $W_r^T\big(W_k^TV_k\big)V_r = I$. The low-rank TBR is then obtained as:
\begin{align}
A_r=W_r^T\big(W_k^TAV_k\big)V_r,\quad B_r=W_r^T\big(W_k^TB\big),\quad C_r=\big(CV_k\big)V_r.\nonumber
\end{align}
It is evident that the low-rank TBR can be seen as a two-step procedure. First, a $k^{th}$-order interpolant $CV_k(sW_k^TV_k - W_k^TAV_k)^{-1}W_k^TB$ is constructed, which interpolates $H(s)$ at $\nu_i$ and $\mu_i$. Then, it is reduced to order $r$ to obtain the low-rank TBR. When $p = m = 1$, $k = r$, and in this case, the low-rank TBR is an interpolant of $H(s)$ at $\nu_i$ and $\mu_i$.

To summarize, the low-rank Krylov-subspace-based method and LRCF-ADI method implicitly perform block interpolation to construct TBR, TCR, and TOR. Consequently, the accuracy of these methods relies on the selection of interpolation points. Typically, these methods interpolate $H(s)$ at multiple interpolation points to capture the majority of its dynamics, and then reduce the resulting interpolant to order $r$ to obtain a compact ROM.
\subsection{Automatic Selection of Interpolation Data and Order}
Up to this point, we have made two main observations. The first is that when the Gramians are low-rank and the Hankel singular values decay rapidly, preserving $r$ significant singular values of $P$, preserving $r$ significant singular values of $Q$, or preserving $r$ significant Hankel singular values of $H(s)$ reduces to tangential interpolation problems. To this end, we have identified the necessary interpolation points and tangential directions required to achieve these goals. The second observation is that low-rank methods for these problems perform block interpolation to address these three cases. Here, the interpolation points are user-defined, and the accuracy of these methods directly depends on the selection of interpolation points made by the user. Readers are referred to \cite{benner2014self} for a detailed discussion on the automatic selection of interpolation points (shifts) in the low-rank ADI method for solving Lyapunov and Sylvester equations. The automatic selection of interpolation data and order in this subsection is based on the findings of subsections \ref{sub1}-\ref{sub3}. 

The appropriate interpolation data for preserving $r$ significant singular values of $P$ is not known beforehand. It only becomes available after constructing the TCR, which inherently preserves the $r$ significant singular values of $P$. Consequently, interpolating at $-\tilde{\lambda}_i$ in the direction of $\tilde{b}_i$ of the TCR to preserve these singular values appears redundant, as they are already preserved by the TCR. Nonetheless, this information about the interpolation data is not entirely redundant. It can be utilized to iteratively refine the interpolation data once an initial arbitrary guess has been made, as will be elaborated later. We begin by presenting our proposed algorithm, referred to as ``Adaptive Low-rank Solver for Lyapunov Equation (ALRS-LYAP),'' and then proceed to explain each step of the algorithm along with the rationale behind it. The pseudo-code for ALRS-LYAP is provided in Algorithm \ref{alg1}.

\begin{algorithm}
\caption{ALRS-LYAP}\label{alg1}
\textbf{Input:} Matrices of Lyapunov Equation: $(A,B)$; Initial order: $r$; Increment in order: $\Delta r$; Tolerance: $tol$; Maximum individual iteration: $i_{max}$ Maximum total iterations: $k_{max}$.\\
\textbf{Output:} Low-rank Approximation of $P$: $\tilde{P}$.
\begin{enumerate}
\item Set $V_k=[$ $]$, $\hat{P}=1$, $V_r=0$, $P_r=0$, $S_r^{(0)}=0$, $k=1$, and $i=1$.
\item Generate arbitrary matrices: $A_r\in\mathbb{R}^{r\times r}$ and $B_r\in\mathbb{R}^{r\times m}$.
\item \textbf{while}$\Big(\frac{S_{r}(r,r)}{S_r(1,1)}>=tol$ and $k\leq k_{max}\Big)$\label{alg1_step3}
\item Solve the Sylvester equation: $A\hat{P}+\hat{P}A_r^T+BB_r^T=0$.\label{alg1_step4}
\item Expand $V_k=\begin{bmatrix}V_k&\hat{P}\end{bmatrix}$ and orthogonalize $V_k=orth(V_k)$.\label{alg1_step5}
\item Solve the Lyapunov equation: $V_k^TAV_kP_k+P_kV_k^TA^TV_k+V_k^TBB^TV_k=0$.\label{alg1_step6}
\item Decompose $P_k$ into $P_k=Z_pZ_p^T$. \label{alg1_step7}
\item Compute SVD of $Z_p^TZ_p$ as $Z_p^TZ_p=\begin{bmatrix}U_r&U_{k-r}\end{bmatrix}\begin{bmatrix}S_r&0\\0&S_{k-r}\end{bmatrix}\begin{bmatrix}U_r^T\\U_{k-r}^T\end{bmatrix}$.\label{alg1_step8}
\item \textbf{if} $\Big(\frac{||S_r^{(i)}-S_r^{(i-1)}||_2}{||S_r^{(i)}||_2}\leq tol$ or $i>=i_{max}\Big)$\label{alg1_step9}
\item Set $r=r+\Delta r$, $V_r=V_kZ_pU_rS_r^{-\frac{1}{2}}$, $V_k=\hat{P}$, $S_r^{(0)}=0$, and $i=0$.\label{alg1_step10}
\item \textbf{else}
\item Set $V_r=V_kZ_pU_rS_r^{-\frac{1}{2}}$.
\item \textbf{end if}
\item Update $A_r=V_r^TAV_r$, $B_r=V_r^TB$, $i=i+1$, and $k=k+1$.\label{alg1_step14}
\item \textbf{end while}
\item Solve the Lyapunov equation: $A_rP_r+P_rA_r^T+B_rB_r^T=0$.
\item Set $\tilde{P}=V_rP_rV_r^T$.\label{st10}
\end{enumerate}
\end{algorithm}
The ALRS-LYAP algorithm starts with an arbitrary pair \((A_r, B_r)\) where \(A_r\) is Hurwitz. The while loop in Step \ref{alg1_step3} keeps increasing the rank of the approximation \(\tilde{P} \approx P\) within the maximum allowed iterations $k_{max}$ until the approximated singular values of \(P\) decay below the tolerance.   In Step \ref{alg1_step4}, reduction matrices are computed to enforce interpolation at the mirror images of the poles of \(A_r\) in the direction of the residuals of \((A_r, B_r)\). The Sylvester equation-based framework from \cite{xu2011optimal} is used to impose tangential interpolation via \(\hat{P}\), which does not require \(A_r\) to have simple poles (see \cite{xu2011optimal} for details). This interpolation data is added to \(V_k\) in Step \ref{alg1_step5}. Each iteration appends \(r\) new columns to \(V_k\), enabling interpolation at \(r\) additional points while preserving prior interpolation conditions. The accuracy of low-rank TCR inherently depends on the interpolant \(CV_k(sI - V_k^T A V_k)^{-1} V_k^T B\), as discussed in Subsection \ref{sub4}. In Step \ref{alg1_step6}, \(V_k\) computes a low-rank approximation of \(P\), which is then used in Steps \ref{alg1_step7}-\ref{alg1_step14} to derive the \(r^{th}\)-order low-rank TCR. Although exact TCR is unavailable, the low-rank TCR acts as a surrogate, updating interpolation data per Subsection \ref{sub1}. Using this, \(r\) interpolation points and tangential directions are generated automatically. As \(V_k\) gains column rank and incorporates more interpolation conditions, the low-rank TCR’s accuracy improves, refining subsequent selections of \(r\) interpolation points and directions. This self-reinforcing process continues within $i_{max}$ iterations until convergence, when the approximated \(r\) singular values of \(P\) in \(S_r\) stagnate. After successfully approximating \(r\) singular values of \(P\), \(r\) is incremented by \(\Delta r\), and \(V_k\) is reset to \(\hat{P}\) (Step \ref{alg1_step10}), which holds refined interpolation data for \(r\) singular values. This reset prevents \(V_k\)’s column rank from growing excessively, keeping the computational cost of SVD in Step \ref{alg1_step8} in check. Even after resetting, \(V_k\) retains refined data from the prior \(r\) interpolation conditions, which are associated with the \(r\) dominant singular values of \(P\). The process repeats until the rank increase results in capturing the insignificant singular values of \(P\), at which point ALRS-LYAP stops, producing \(P \approx \tilde{P} = V_r P_r V_r^T\).  ALRS-LYAP is fully automatic and adaptive in selecting interpolation data and rank \(r\) of \(\tilde{P}\). The user only need to specify: (i) Error tolerance (\(tol\)), (ii) Starting rank ($r$), (iii) Rank increment (\(\Delta r\)), (iv) Convergence wait limits ($k_{max}$, $i_{max}$). Based on these, ALRS-LYAP adaptively chooses tangential interpolation data guided by Subsection \ref{sub1}, requiring no further input.

In the literature, low-rank TBR is typically computed by treating the low-rank approximations of $P$ and $Q$ as two independent problems. If we were to follow this conventional approach, we could use ALRS-LYAP to compute the low-rank approximations of $P$ and $Q$, and subsequently apply the balancing square-root algorithm to derive the low-rank TBR. However, a significant drawback of this approach is that a state that is poorly controllable but strongly observable might still have a significant Hankel singular value associated with it. ALRS-LYAP would not specifically target such a state, and it might not be captured in the approximation $\tilde{P}$. Similarly, a state that is poorly observable but strongly controllable, and thus associated with significant Hankel singular values, might not be captured in the approximation of $Q$. Let us examine a small illustrative example that highlights the limitations of the approach predominantly used in the literature.

\textbf{Illustrative Example:} Consider a fourth-order model represented by the following state-space realization: 
\begin{align}
A=\begin{bmatrix}-0.1 &0& 0& 0\\ 0& -0.2& 0& 0\\ 0 &0& -100& 0\\ 0& 0& 0& -200\end{bmatrix},\quad B=\begin{bmatrix}1\\1\\10^4\\1\end{bmatrix},\quad C=\begin{bmatrix}1&1&1&10^4\end{bmatrix}.\nonumber
\end{align}Since this realization is in modal form, the controllability and observability of each pole, along with their respective Hankel singular values, can be assessed through visual inspection of the poles and residuals; refer to \cite{gawronski2004dynamics} for an overview of the modal approach in system dynamics analysis. The pole at $-100$ exhibits strong controllability due to its large input residual ($10^4$), but it is weakly observable since it is located far from the $j\omega$-axis in the $s$-plane. Conversely, the pole at $-200$ is highly observable due to its large output residual ($10^4$), but it is weakly controllable as it is located far from the $j\omega$-axis. The singular values of $P$ are given as $5\times10^5$, $7.2713$, $0.1887$, and $0.0002$, with the smallest singular value corresponding to the pole at $-200$. Since the fourth singular value is negligible, we can truncate its singular value decomposition to obtain a rank-$3$ approximation $\tilde{P}$, with a relative error of $\frac{||P-\tilde{P}||_2}{||P||_2}=5.5223\times 10^{-10}$. In the literature, low-rank approximation accuracy is often assessed via the relative residual $\frac{||A\tilde{P}+\tilde{P}A^T+BB^T||_2}{||BB^T||_2}$, which, in this case, is $1.1045\times 10^{-9}$. Although residuals may sometimes be misleading in determining accuracy, both indicators confirm that the rank-$3$ approximation provides a good approximation of $P$. Similarly, the singular values of $Q$ are $2.5\times10^5$, $7.2906$, $0.18936$, and $0.0005$, with the smallest singular value associated with the pole at $-100$. By truncating its singular value decomposition, we obtain a rank-$3$ approximation $\tilde{Q}$ with a relative error of $\frac{||Q-\tilde{Q}||_2}{||Q||_2}=2.1957\times10^{-9}$ and a relative residual of $\frac{||A^T\tilde{Q}+\tilde{Q}A+C^TC||_2}{||C^TC||_2}=1.0978\times 10^{-9}$. Both indicators verify that the rank-$3$ approximation accurately represents $Q$. Given these excellent approximations of $P$ and $Q$, one might expect that using $\tilde{P}=Z_pZ_p^T$ and $\tilde{Q}=Z_qZ_q^T$ in the balancing square-root algorithm would preserve the original system’s Hankel singular values, which are $73.1370$, $7.2831$, $1.8919$, and $0.1880$. However, the Hankel singular values of the second-order reduced model using these low-rank approximations are $72.9579$ and $8.3810$. The most significant Hankel singular value ($73.1370$) is associated with the pole at $-100$, whose observability information was truncated in the approximation $\tilde{Q}$.
The second most significant Hankel singular value ($7.2831$) is linked to the pole at $-200$, whose controllability information was lost in the approximation $\tilde{P}$. As a result, the low-rank BT method relying on $\tilde{P}$ and $\tilde{Q}$ failed to preserve the two most significant Hankel singular values, despite both approximations being excellent representations of $P$ and $Q$. ALRS-LYAP is more likely to produce a good low-rank TBR only when the states that are poorly controllable are also poorly observable. For this reason, we avoid treating $P$ and $Q$ as independent approximation problems. Instead, we build upon the results of subsection \ref{sub3} to develop our low-rank algorithm for BT. 

The appropriate interpolation data for preserving the $r$ significant Hankel singular values of $H(s)$ is not known in advance and is only determined after constructing the TBR. However, this information is not redundant, as it can serve to iteratively refine the interpolation data once an initial arbitrary guess is provided, as will be explained shortly. Again, we present our proposed algorithm, referred to as ``Adaptive Tangential Interpolation Algorithm for Balanced Truncation (ATIA-BT),'' first. We then proceed to explain each step of the algorithm and the rationale behind it in detail. The pseudo-code for ATIA-BT is provided in Algorithm \ref{alg2}.

\begin{algorithm}
\caption{ATIA-BT}\label{alg2}
\textbf{Input:} State-space matrices: $(A,B,C)$; Initial order: $r$; Increment in order: $\Delta r$; Tolerance: $tol$; Maximum individual iterations: $i_{max}$; Maximum total iterations $k_{max}$.\\
\textbf{Output:} ROM: $(A_r,B_r,C_r)$.
\begin{enumerate}
\item Set $V_k=[$ $]$, $W_k=[$ $]$, $\hat{P}=1$, $\hat{Q}=1$, $P_r=0$, $Q_r=0$, $S_r^{(0)}=0$, $i=1$, and $k=1$.
\item Generate an arbitrary $r^{th}$-order state-space model: $(A_r,B_r,C_r)$.
\item \textbf{while}$\Big(\frac{S_{r}(r,r)}{S_r(1,1)}>=tol$ and $k\leq k_{max}\Big)$
\item Solve the Sylvester equations: $A\hat{P}+\hat{P}A_r^T+BB_r^T=0$,\\
\hspace*{4.5cm}$A^T\hat{Q}+\hat{Q}A_r+C^TC_r=0$.\label{alg2_step4}
\item Expand $V_k=\begin{bmatrix}V_k&\hat{P}\end{bmatrix}$ and $W_k=\begin{bmatrix}W_k&\hat{Q}\end{bmatrix}$.\label{alg2_step11}
\item  Orthogonalize $V_k=orth(V_k)$ and $W_k=orth(W_k)$.
\item  Solve the Lyapunov equations:\\
\hspace*{2cm}$V_k^TAV_kP_k+P_kV_k^TA^TV_k+V_k^TBB^TV_k=0$,\\
\hspace*{2cm}$W_k^TA^TW_kQ_k+Q_kW_k^TAW_k+W_k^TC^TCW_k=0.$\label{alg2_step13}
\item  Decompose $P_k$ and $Q_k$ into $P_k=Z_pZ_p^T$ and $Q_k=Z_qZ_q^T$.\label{alg2_step14}
\item  Compute SVD of $Z_q^TW_k^TV_kZ_p$ as\\
    \hspace*{2cm}$Z_p^TW_k^TV_kZ_p=\begin{bmatrix}U_r&U_{k-r}\end{bmatrix}\begin{bmatrix}S_r&0\\0&S_{k-r}\end{bmatrix}\begin{bmatrix}R_r^T\\R_{k-r}^T\end{bmatrix}$.\label{alg2_step15}
\item \textbf{if} $\Big(\frac{||S_r^{(i)}-S_r^{(i-1)}||_2}{||S_r^{(i)}||_2}$ or $i>=i_{max}\Big)$
\item Set $r=r+\Delta r$, $V_r=V_kZ_pR_rS_r^{-\frac{1}{2}}$, $W_r=W_kZ_qU_rS_r^{-\frac{1}{2}}$, $V_k=\hat{P}$, $W_k=\hat{Q}$, $S_r^{(0)}=0$, and $i=0$.\label{alg2_step17}
\item \textbf{else}
\item  Set $V_r=V_kZ_pR_rS_r^{-\frac{1}{2}}$ and $W_r=W_kZ_qU_rS_r^{-\frac{1}{2}}$.
\item \textbf{end if}
\item Update $A_r=W_r^TAV_r$, $B_r=W_r^TB$, $C_r=CV_r$, $i=i+1$, and $k=k+1$.\label{alg2_step21}
\item  \textbf{end while}
\end{enumerate}
\end{algorithm}
ATIA-BT begins with an arbitrary initial guess \((A_r, B_r, C_r)\), where \(A_r\) is Hurwitz. In Step \ref{alg2_step4}, reduction matrices are computed to enforce bi-tangential Hermite interpolation at the mirror images of the poles of \(A_r\) in the direction of the residuals of \((A_r, B_r, C_r)\). This interpolatory data is then added to \(V_k\) and \(W_k\) in Step \ref{alg2_step11}. At each iteration, \(r\) new columns are appended to \(V_k\) and \(W_k\), enabling Hermite interpolation of \(H(s)\) at \(r\) additional points while keeping prior interpolation conditions. The accuracy of low-rank BT depends on the approximation quality of the interpolant \( CV_k(sW_k^TV_k - W_k^T A V_k)^{-1} W_k^T B \), as detailed in Subsection \ref{sub4}. In Step \ref{alg2_step13}, low-rank approximations of \(P\) and \(Q\) are computed using \(V_k\) and \(W_k\), which are then used to compute the \(r^{th}\)-order low-rank TBR in Steps \ref{alg2_step14}–\ref{alg2_step21}. Though the exact TBR is unknown, the low-rank surrogate—guided by the results in Subsection \ref{sub3}—automatically generates \(r\) interpolation points and \(2r\) tangential directions without user intervention. As \(V_k\) and \(W_k\) accumulate columns and enforce more interpolation conditions, the low-rank TBR improves, refining subsequent selections of interpolation data. This self-correcting process continues within the iteration limit $i_{max}$ until convergence, marked by stagnation in the approximated \(r\) Hankel singular values (stored in \(S_r\)). At this stage (Step \ref{alg2_step17}), the order \(r\) is incremented by \(\Delta r\), and \(V_k\) and \(W_k\) are reset to \(\hat{P}\) and \(\hat{Q}\), respectively—a measure to control SVD costs in Step \ref{alg2_step15}. Despite the reset, \(V_k\) and \(W_k\) retain refined interpolation data from the prior \(r\) Hermite interpolation conditions, tied to the \(r\) dominant Hankel singular values of \(H(s)\). The process repeats within $k_{max}$ iterations until the order increment results in capturing insignificant Hankel singular values, at which point ATIA-BT stops. ATIA-BT is fully automatic and adaptive in selecting interpolation data and the order \(r\) of \(H(s)\). The user need only specify: (i) Error tolerance (\(tol\)), (ii) Starting order (\(r\)), (iii) Order increment (\(\Delta r\)), and (iv) Convergence wait limits ($k_{max}, i_{max}$). Based on these inputs, ATIA-BT adaptively selects tangential interpolation data, guided by the results of Subsection \ref{sub3}, requiring no further user intervention. 
\subsection{Computational Aspects}
Convergence in the ALRS-LYAP and ATIA-BT algorithms is not guaranteed. The accuracy of these algorithms depends on their respective interpolants, which are reduced to obtain the ROM. High-quality interpolation data can significantly improve the likelihood of rapid convergence. As suggested in \cite{gugercin2008h_2}, the initial guess should preferably include the most controllable and most observable poles of $H(s)$, along with their associated residuals. Interpolating at the mirror images of these poles in the direction of their respective residuals ensures a small $\mathcal{H}_2$-norm error from the outset. This initial guess can be efficiently computed using the algorithm proposed in \cite{rommes2006efficient}.  

The main computational burden in ALRS-LYAP and ATIA-BT is concentrated in two key steps: solving Sylvester equations and computing the SVD. All other operations are small-scale and can be performed efficiently. The Sylvester equations in these algorithms belong to a specific category of skinny-tall Sylvester equations, commonly encountered in $\mathcal{H}_2$-optimal MOR algorithms. An efficient method for solving such equations is presented in \cite{MPIMD11-11}. Using this approach, ALRS-LYAP requires solving $r$ linear systems of equations ($Ax=b$) per iteration, while ATIA-BT requires solving $2r$ linear systems of the same form during each iteration. Since $r \ll n$, the computational cost of solving these Sylvester equations remains manageable even for large-scale systems. The computational cost of SVD also remains feasible as the column ranks $k$ of $V_k$ and $W_k$ are kept small. In summary, both ALRS-LYAP and ATIA-BT are capable of handling large-scale systems. The next section demonstrates their computational efficiency by applying them to a large-scale model of the order \(10^7\).
\section{Numerical Results}
This section evaluates the performance of ALRS-LYAP and ATIA-BT using widely used benchmark problems for testing MOR algorithms \cite{chahlaoui2005benchmark}. The first three examples deal with models of modest orders, allowing full-rank computations for \( P \) and \( Q \) and enabling error analysis. In contrast, the fourth example is a large-scale problem where full-rank computation of \( P \) and \( Q \) is infeasible. This example demonstrates the computational efficiency of ALRS-LYAP and ATIA-BT. All simulations are executed in MATLAB R2021b on a Windows 11 laptop equipped with a 2GHz Intel Core i7 processor and 16GB of RAM.

\textbf{Example 1: CD Player (Continued)}\\
We revisit the CD player model from the previous section. In this example, the initial order \( r \) is set to 2, with an increment \( \Delta r \) of 2, i.e., \( r = 2 \) and \( \Delta r = 2 \). The maximum allowable iterations are set as $i_{max} = 5$ and $k_{max} = 35$. These settings remain fixed for all experiments on the CD player model. An arbitrary state-space model of order 2 is generated using MATLAB's \textit{rss} command, which is also kept constant throughout the experiments. The tolerance \( tol \) is then varied from \( 10^{-4} \) to \( 10^{-6} \). The accuracy of the low-rank approximations of \( P \) and \( Q \) obtained by varying \( tol \) in ALRS-LYAP is presented in Table \ref{tab1}.

\begin{table}[!h]
\centering
\caption{Accuracy of the low-rank approximations of $P$ and $Q$}\label{tab1}
\begin{tabular}{|c|c|c|}\hline
$tol$&$\frac{||P-V_rP_rV_r^T||_2}{||P||_2}$&$\frac{||Q-W_rQ_rW_r^T||_2}{||Q||_2}$\\\hline
$10^{-4}$&$0.0015$ ($r=6$)&$0.0015$ ($r=6$)\\
$10^{-5}$&$1.1172\times10^{-5}$ ($r=10$)&$3.5702\times10^{-6}$ ($r=12$)\\
$10^{-6}$&$5.4863\times10^{-7}$ ($r=16$)&$9.6199\times10^{-7}$ ($r=18$)\\\hline
\end{tabular}
\end{table}

\begin{figure}[!h]
  \centering
  \includegraphics[width=8cm]{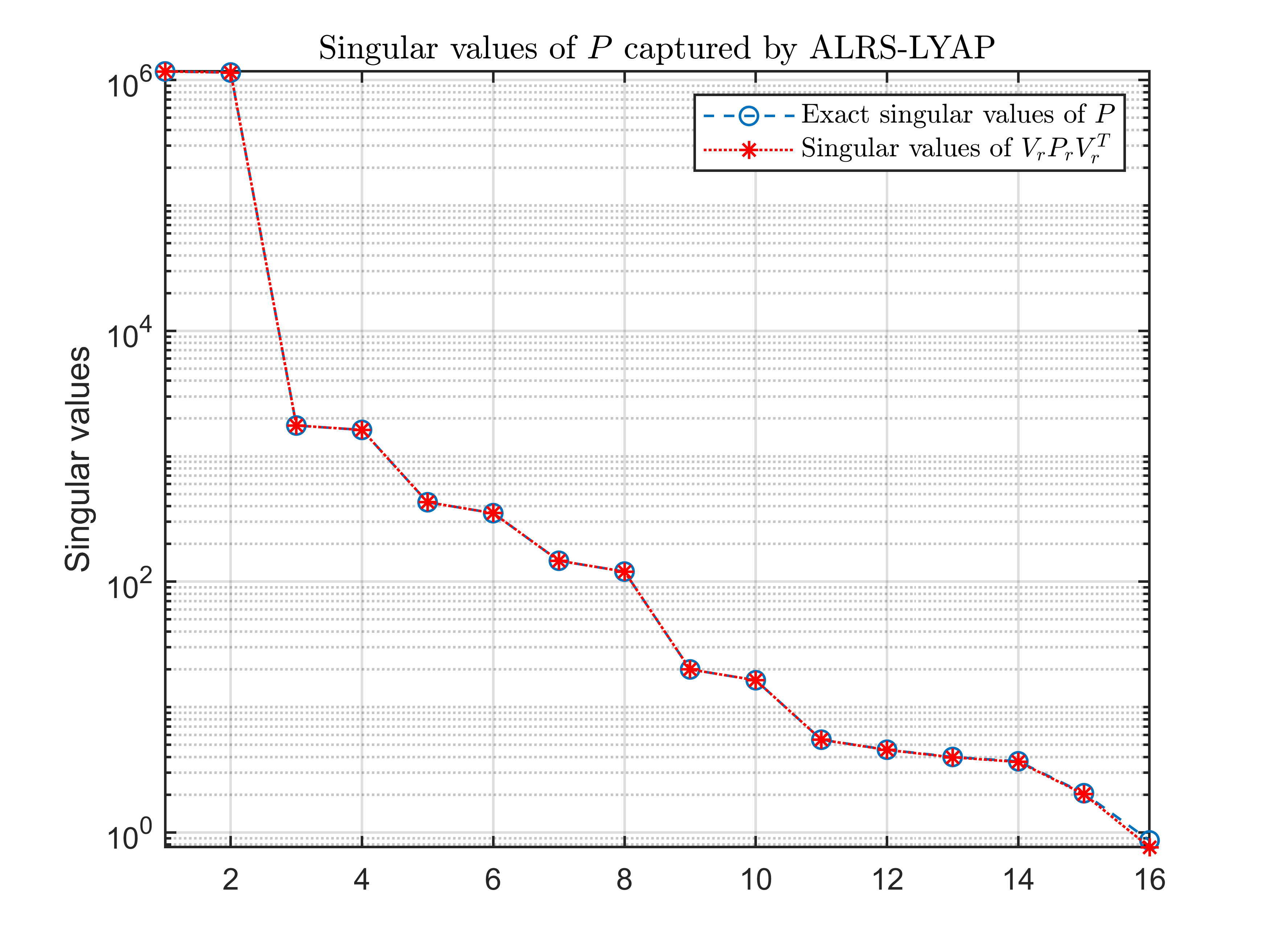}
  \caption{Singular values of $P$ and $V_rP_rV_r^T$}\label{fig04}
\end{figure}

\begin{figure}[!h]
  \centering
  \includegraphics[width=8cm]{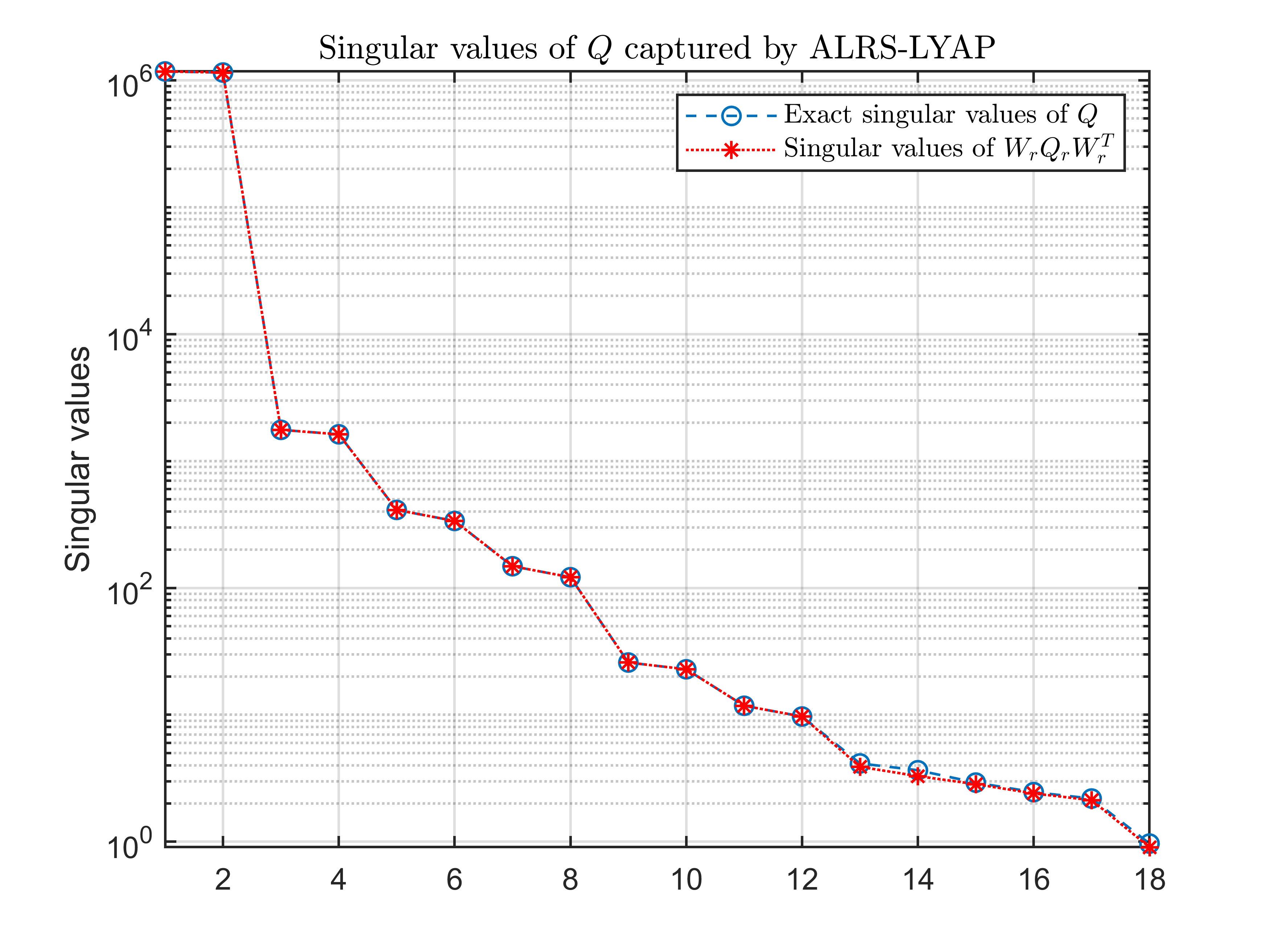}
  \caption{Singular values of $Q$ and $W_rQ_rW_r^T$}\label{fig05}
\end{figure}

As the tolerance \( tol \) decreases, ALRS-LYAP's accuracy improves, as indicated by the reduction in the relative errors \( \frac{\| P - V_r P_r V_r^T \|_2}{\| P \|_2} \) and \( \frac{\| Q - W_r Q_r W_r^T \|_2}{\| Q \|_2} \). The $16$ largest singular values of \( P \) and the $18$ largest singular values of \( Q \) captured by ALRS-LYAP, with \( tol \) set to \( 10^{-6} \), are compared against the exact singular values of \( P \) and \( Q \) in Figure \ref{fig04} and Figure \ref{fig05}, respectively. As seen in these figures, ALRS-LYAP successfully captures the largest singular values of \( P \) and \( Q \).
It is interesting to note that the truncated singular values of \( P \) and \( Q \) are not near-zero, contradicting the assumptions in Subsections \ref{sub1} and \ref{sub2}. However, this did not prevent ALRS-LYAP from accurately capturing the dominant singular values of \( P \) and \( Q \), consistent with the observations made in the previous section where this example was initially examined.

\begin{figure}[!h]
  \centering
  \includegraphics[width=8cm]{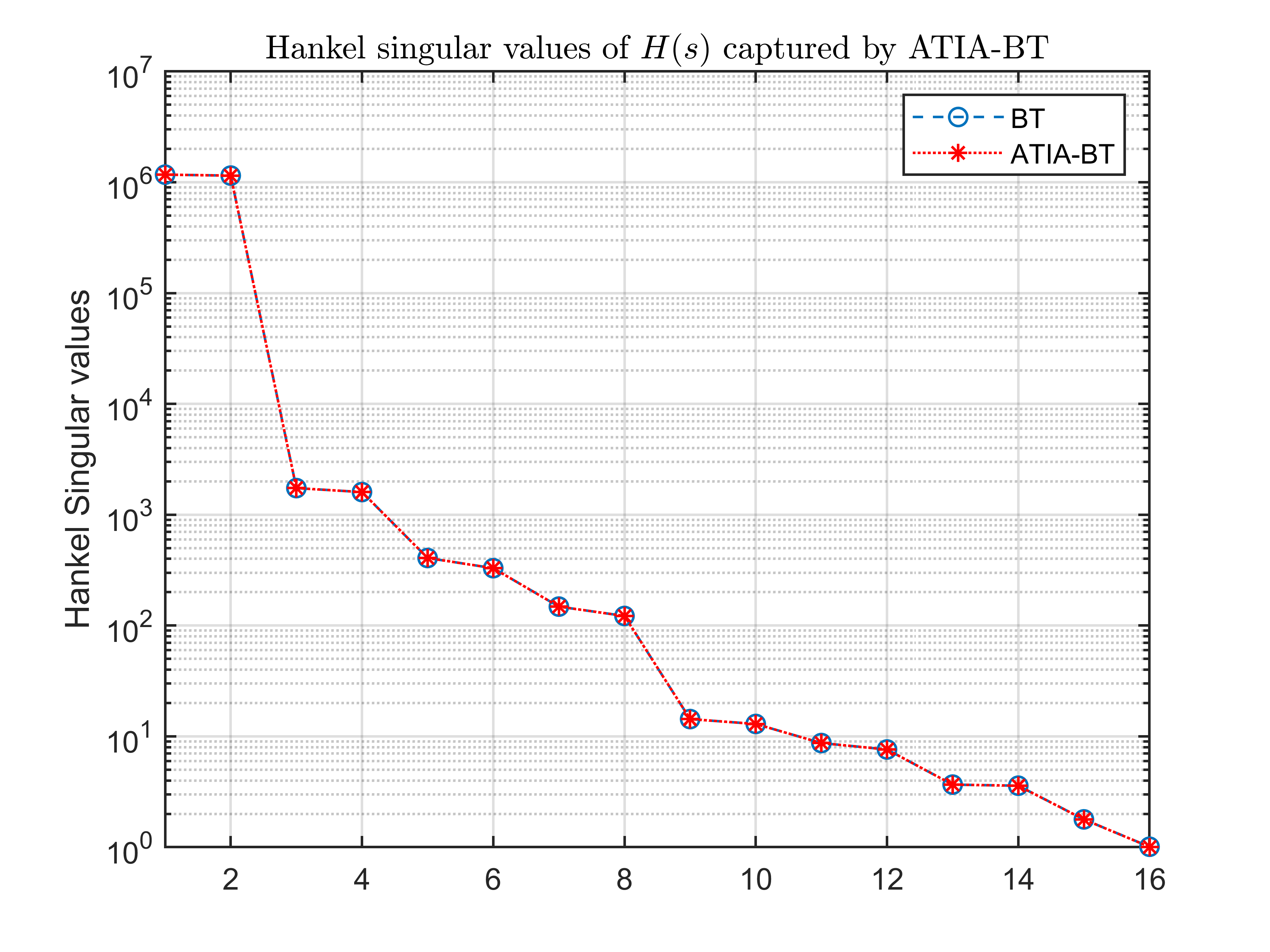}
  \caption{Hankel Singular Values of $16^{th}$-order $H_r(s)$}\label{fig06}
\end{figure}

\begin{figure}[!h]
  \centering
  \includegraphics[width=8cm]{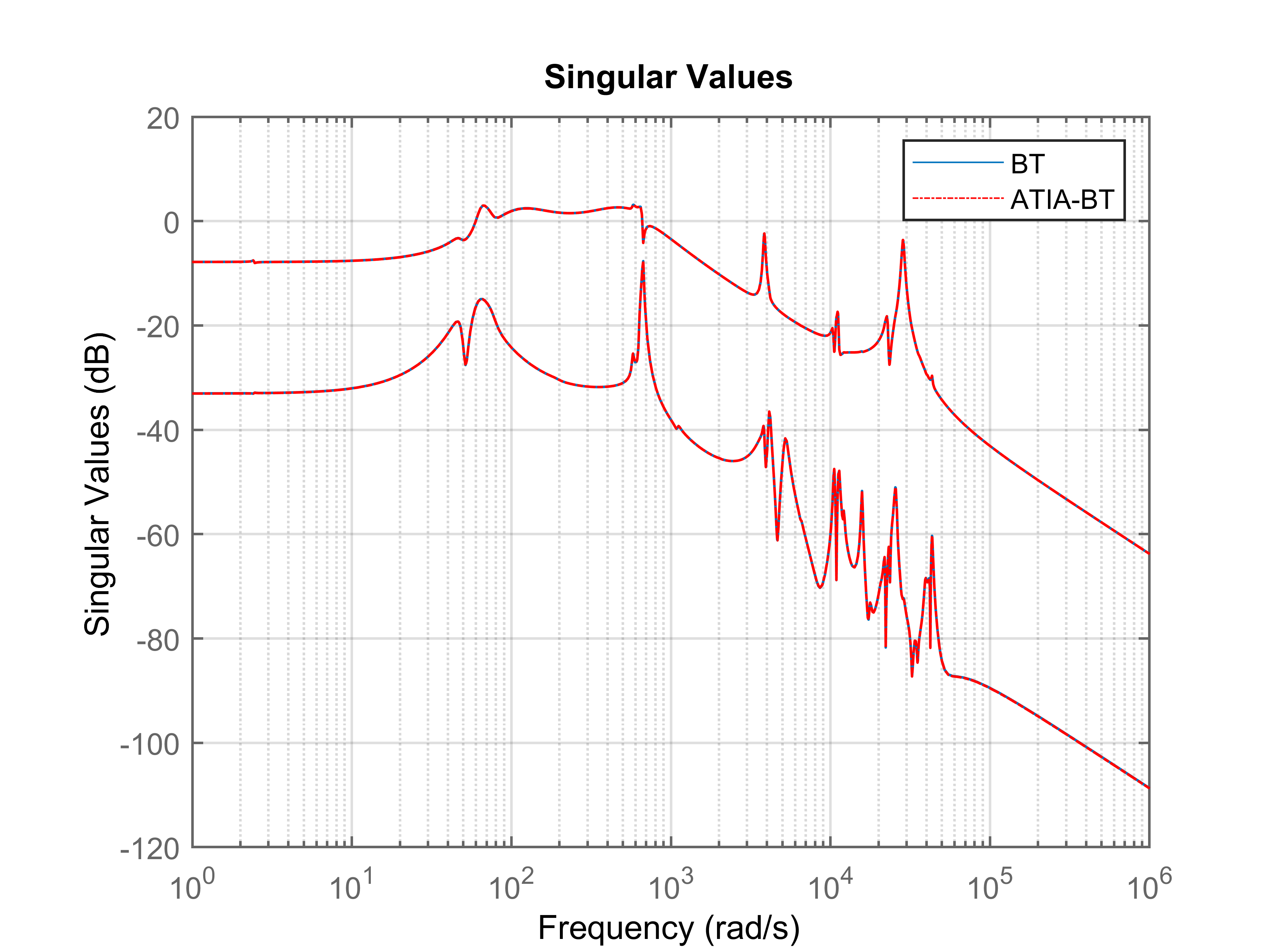}
  \caption{Singular values of $H(s)-H_r(s)$}\label{fig07}
\end{figure}

To evaluate the performance of ATIA-BT, the tolerance \( tol \) is varied from \( 10^{-4} \) to \( 10^{-6} \), and the relative error \( \frac{\| G(s) - G_r(s) \|_{\mathcal{H}_\infty}}{\| G(s) \|_{\mathcal{H}_\infty}} \) is presented in Table \ref{tab2}. As \( tol \) decreases, the relative error falls almost identically to that of BT. The Hankel singular values of the $16^{th}$-order ROM produced by ATIA-BT with \( tol = 10^{-6} \), along with those from BT, are plotted in Figure \ref{fig06}. As shown, ATIA-BT successfully captures the $16$ largest Hankel singular values of \( H(s) \), similar to BT. Additionally, Figure \ref{fig07} displays the singular values of \( H(s) - H_r(s) \) for the $16^{th}$-order ROMs obtained from both methods, confirming that ATIA-BT and BT achieve nearly identical accuracy.
\begin{table}[!h]
\centering
\caption{Comparison of Relative Error $\frac{||G(s)-G_r(s)||_{\mathcal{H}_\infty}}{||G(s)||_{\mathcal{H}_\infty}}$}\label{tab2}
\begin{tabular}{|c|c|c|c|}\hline
$tol$&r&ATIA-BT&BT\\\hline
$10^{-4}$&6&$0.0015$&$1.2014\times10^{-4}$\\
$10^{-5}$&12&$2.7468\times10^{-6}$&$2.7479\times10^{-6}$\\
$10^{-6}$&16&$6.1721\times10^{-7}$&$6.1833\times10^{-7}$\\\hline
\end{tabular}
\end{table}

\textbf{Example 2: Artificial Dynamical System}\\
The artificial dynamical system, taken from the benchmark collection in \cite{chahlaoui2005benchmark}, is a $1006^{th}$-order single-input single-output system extensively used for testing MOR algorithms in the literature. In this example, the initial order \( r \) is set to 2, with an increment \( \Delta r \) of 2, i.e., \( r = 2 \) and \( \Delta r = 2 \). The maximum allowable iterations are fixed at $i_{max} = 5$ and $k_{max} = 35$ for all experiments conducted on this model. A state-space model of order $2$ is generated using MATLAB’s \textit{rss} command and remains unchanged throughout the experiments. The tolerance \( tol \) is varied from \( 10^{-4} \) to \( 10^{-6} \), and the accuracy of the low-rank approximations of \( P \) and \( Q \) is shown in Table \ref{tab3}. 
\begin{table}[!h]
\centering
\caption{Accuracy of the low-rank approximations of $P$ and $Q$}\label{tab3}
\begin{tabular}{|c|c|c|}\hline
$tol$&$\frac{||P-V_rP_rV_r^T||_2}{||P||_2}$&$\frac{||Q-W_rQ_rW_r^T||_2}{||Q||_2}$\\\hline
$10^{-4}$&$5.1527\times10^{-6}$ ($r=14$)&$5.1527\times10^{-6}$ ($r=14$)\\
$10^{-5}$&$3.9555\times10^{-7}$ ($r=16$)&$3.9555\times10^{-7}$ ($r=16$)\\
$10^{-6}$&$2.8336\times10^{-8}$ ($r=18$)&$2.8336\times10^{-8}$ ($r=18$)\\\hline
\end{tabular}
\end{table}

\begin{figure}[!h]
  \centering
  \includegraphics[width=8cm]{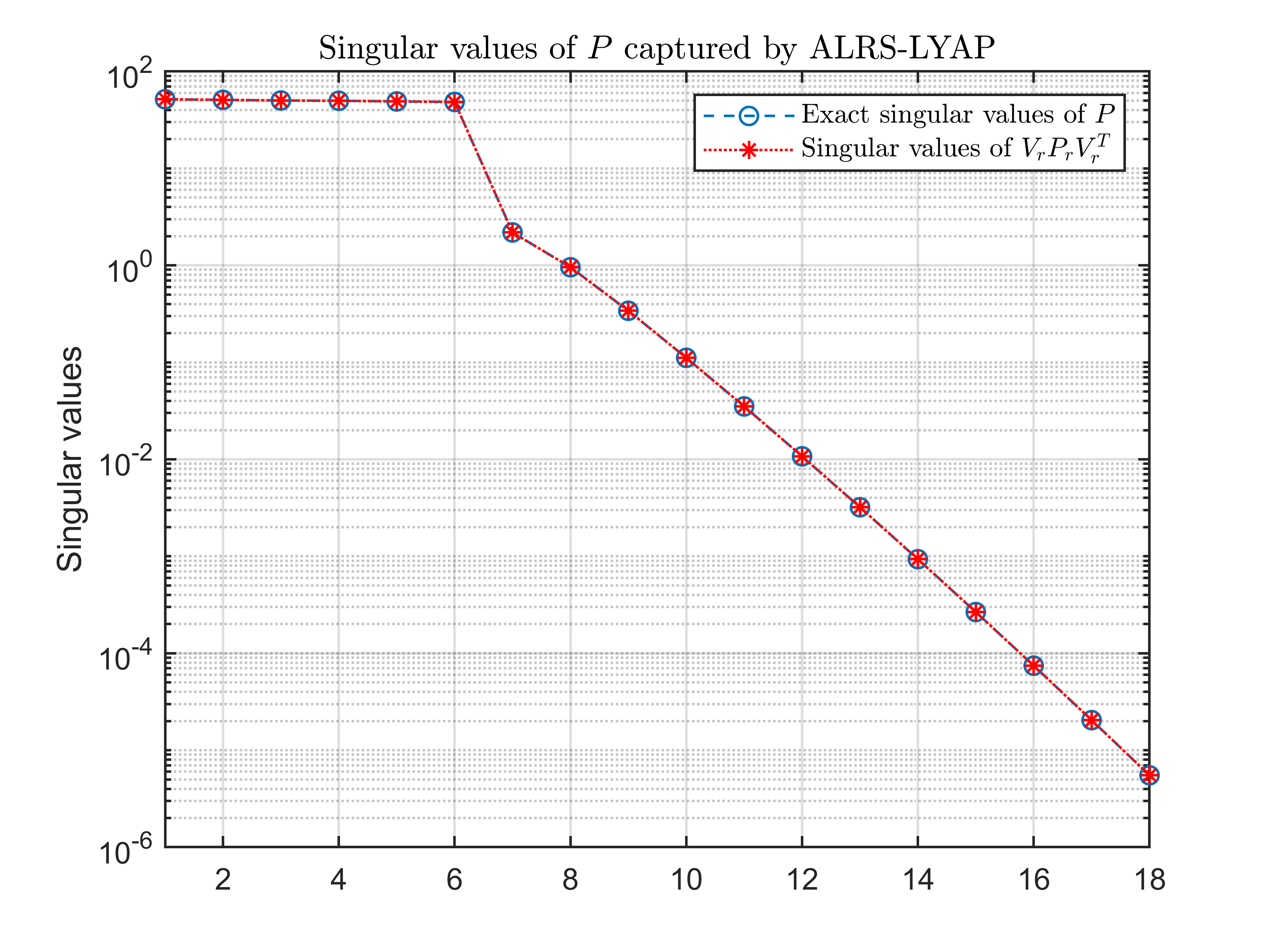}
  \caption{Singular values of $P$ and $V_rP_rV_r^T$}\label{fig08}
\end{figure}

\begin{figure}[!h]
  \centering
  \includegraphics[width=8cm]{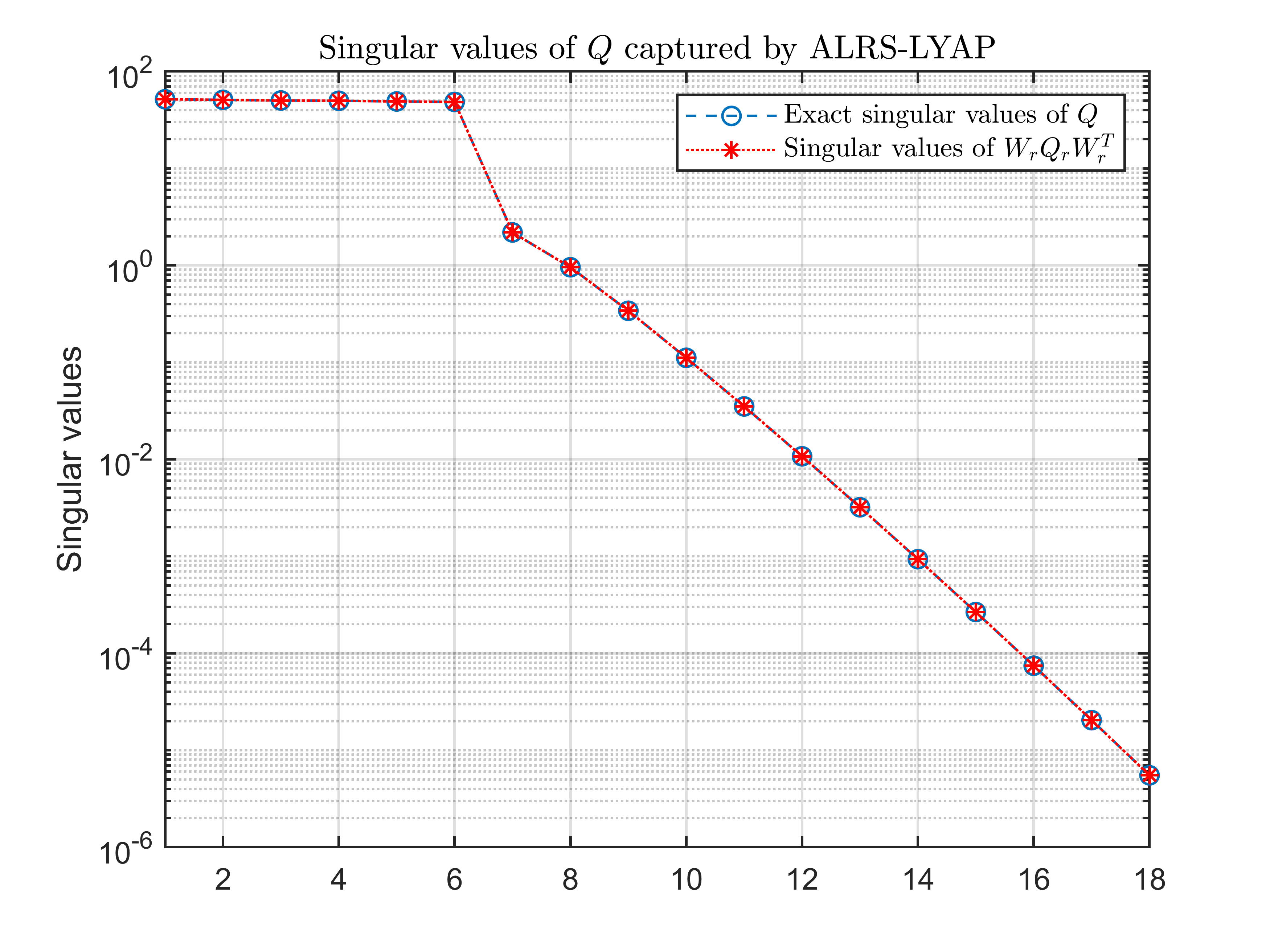}
  \caption{Singular values of $Q$ and $W_rQ_rW_r^T$}\label{fig09}
\end{figure}

As \( tol \) decreases, ALRS-LYAP's accuracy improves, reflected in the reduction of the relative errors \( \frac{\| P - V_r P_r V_r^T \|_2}{\| P \|_2} \) and \( \frac{\| Q - W_r Q_r W_r^T \|_2}{\| Q \|_2} \). The $18$ largest singular values of \( P \) and the $18$ largest singular values of \( Q \) captured by ALRS-LYAP, with \( tol = 10^{-6} \), are compared to the exact singular values of \( P \) and \( Q \) in Figure \ref{fig08} and Figure \ref{fig09}, confirming that ALRS-LYAP effectively captures the dominant singular values.

\begin{figure}[!h]
  \centering
  \includegraphics[width=8cm]{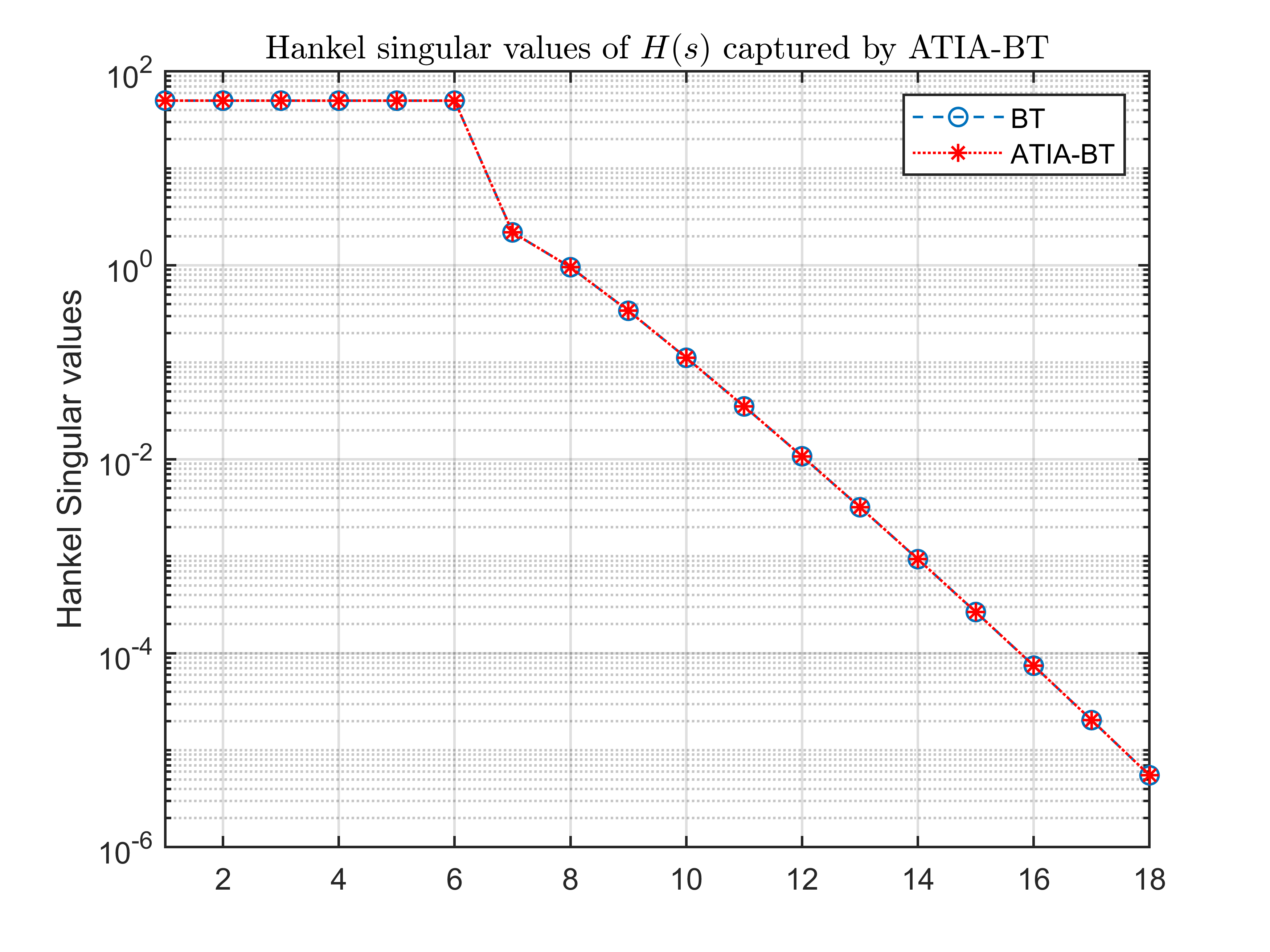}
  \caption{Hankel Singular Values of $18^{th}$-order $H_r(s)$}\label{fig10}
\end{figure}

\begin{figure}[!h]
  \centering
  \includegraphics[width=8cm]{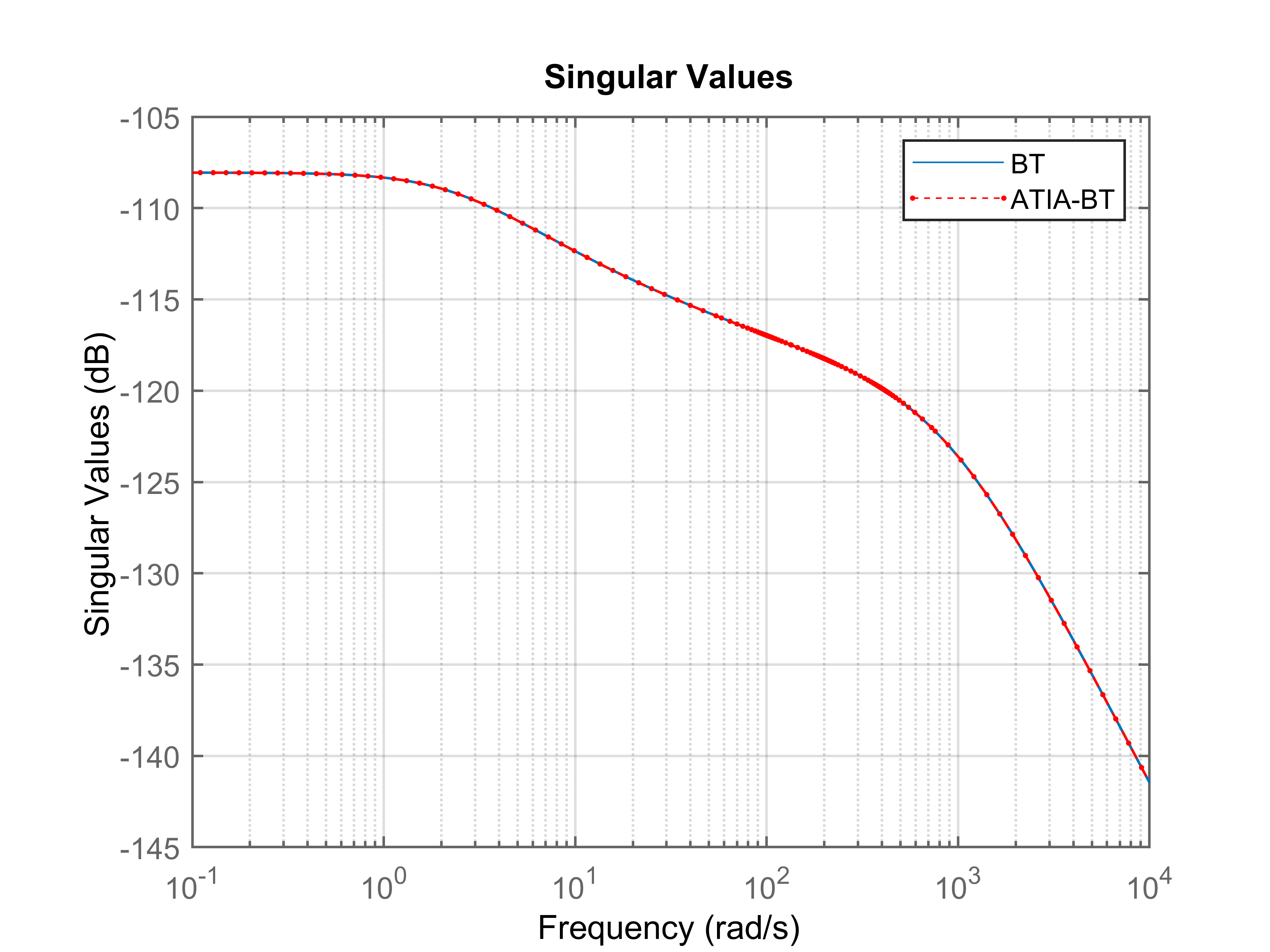}
  \caption{Singular values of $H(s)-H_r(s)$}\label{fig11}
\end{figure}

To assess ATIA-BT’s performance, \( tol \) is varied from \( 10^{-4} \) to \( 10^{-6} \), and the relative error \( \frac{\| G(s) - G_r(s) \|_{\mathcal{H}_\infty}}{\| G(s) \|_{\mathcal{H}_\infty}} \) is tabulated in Table \ref{tab4}. As \( tol \) decreases, the relative error falls almost identically to that of BT. The Hankel singular values of the $18^{th}$-order ROM produced by ATIA-BT with \( tol = 10^{-6} \) are compared with those from BT in Figure \ref{fig10}, showing that ATIA-BT successfully captured the $18$ largest Hankel singular values of \( H(s) \). Additionally, Figure \ref{fig11} plots the singular values of \( H(s) - H_r(s) \) for the $18^{th}$-order ROMs obtained by both methods, demonstrating that ATIA-BT and BT yield identical accuracy.
\begin{table}[!h]
\centering
\caption{Comparison of Relative Error $\frac{||G(s)-G_r(s)||_{\mathcal{H}_\infty}}{||G(s)||_{\mathcal{H}_\infty}}$}\label{tab4}
\begin{tabular}{|c|c|c|c|}\hline
$tol$&r&ATIA-BT&BT\\\hline
$10^{-4}$&14&$7.2086\times 10^{-6}$&$7.1996\times 10^{-6}$\\
$10^{-5}$&16&$5.4560\times 10^{-7}$&$5.4560\times 10^{-7}$\\
$10^{-6}$&18&$3.8651\times 10^{-8}$&$3.8651\times 10^{-8}$\\\hline
\end{tabular}
\end{table}

\textbf{Example 3: International Space Station}\\
The international space station model, taken from the benchmark collection in \cite{chahlaoui2005benchmark}, is a $270^{th}$-order system with $3$ inputs and $3$ outputs, widely used for testing MOR algorithms in the literature. In this example, the initial order \( r \) is set to $5$ with an increment \( \Delta r \) of $5$, i.e., \( r = 5 \) and \( \Delta r = 5 \). The maximum permissible iterations are set at $i_{max} = 5$ and $k_{max} = 45$ across all experiments on this model. A state-space model of order $5$ is generated using MATLAB’s \textit{rss} command and remains unchanged throughout the experiments. The tolerance \( tol \) is varied from \( 10^{-4} \) to \( 10^{-6} \), with the accuracy of low-rank approximations of \( P \) and \( Q \) presented in Table \ref{tab5}.  
\begin{table}[!h]
\centering
\caption{Accuracy of the low-rank approximations of $P$ and $Q$}\label{tab5}
\begin{tabular}{|c|c|c|}\hline
$tol$&$\frac{||P-V_rP_rV_r^T||_2}{||P||_2}$&$\frac{||Q-W_rQ_rW_r^T||_2}{||Q||_2}$\\\hline
$10^{-4}$&$1.0167\times 10^{-4}$ ($r=50$)&$0.0818$ ($r=25$)\\
$10^{-5}$&$1.0167\times 10^{-4}$ ($r=50$)&$6.8621\times10^{-5}$ ($r=50$)\\
$10^{-6}$&$1.0167\times 10^{-4}$ ($r=50$)&$6.8621\times10^{-5}$ ($r=50$)\\\hline
\end{tabular}
\end{table}

\begin{figure}[!h]
  \centering
  \includegraphics[width=8cm]{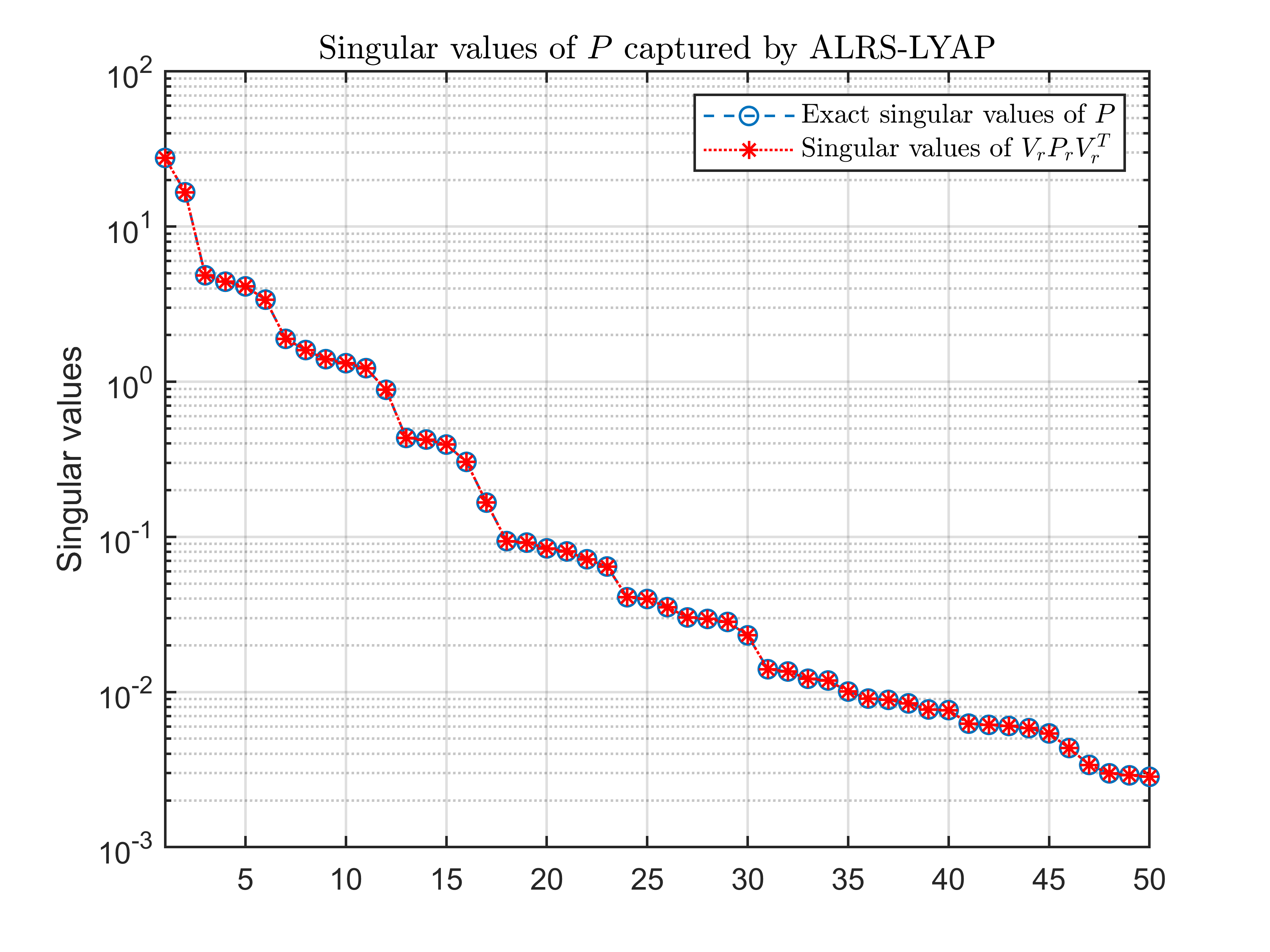}
  \caption{Singular values of $P$ and $V_rP_rV_r^T$}\label{fig12}
\end{figure}

\begin{figure}[!h]
  \centering
  \includegraphics[width=8cm]{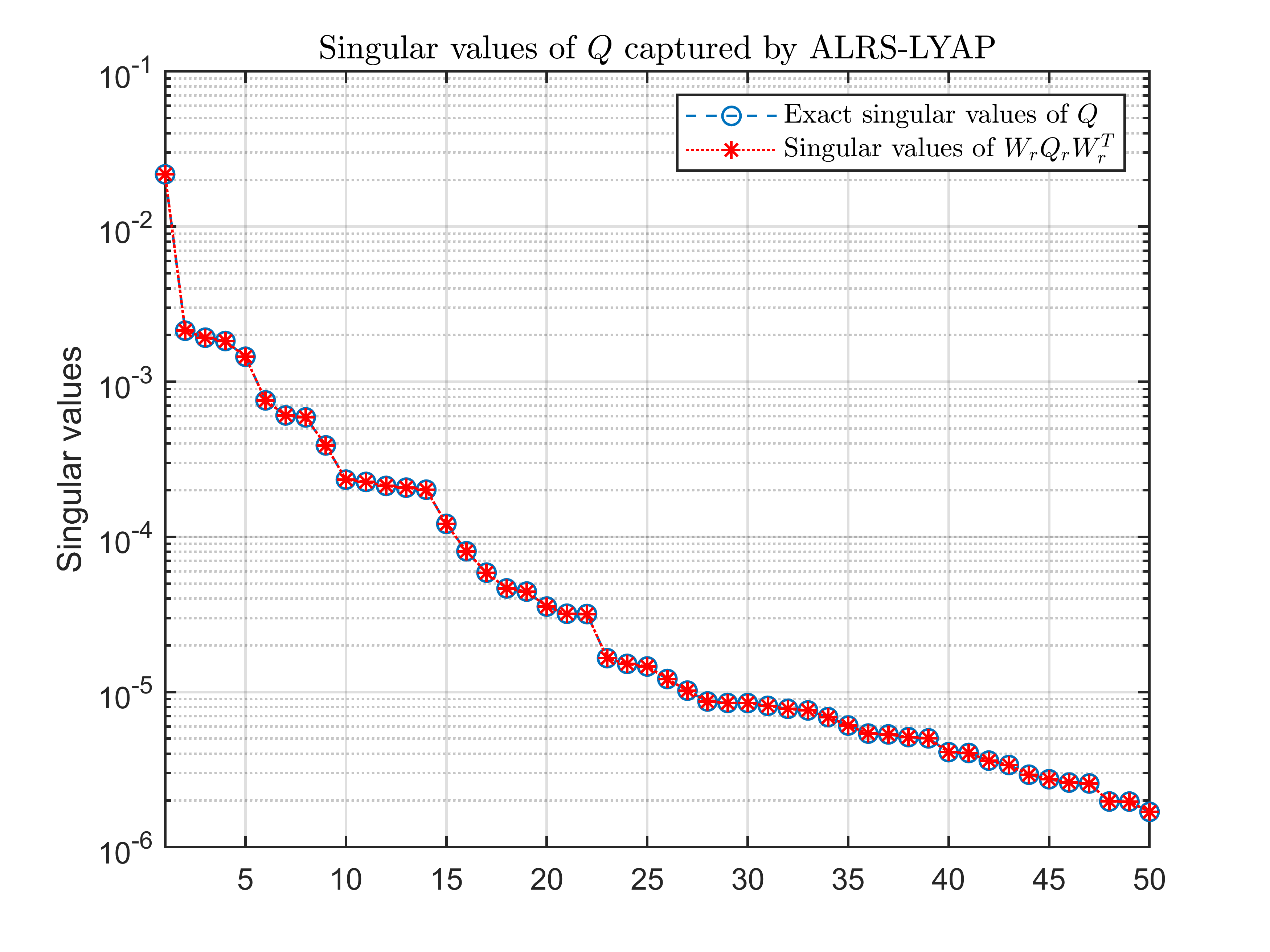}
  \caption{Singular values of $Q$ and $W_rQ_rW_r^T$}\label{fig13}
\end{figure}

In this example, ALRS-LYAP failed to converge, and adjusting the tolerance \( tol \) had no impact on accuracy. The $50$ largest singular values of \( P \) and \( Q \) captured by ALRS-LYAP with \( tol = 10^{-4} \) are compared with the exact singular values of \( P \) and \( Q \) in Figure \ref{fig12} and Figure \ref{fig13}, confirming its effectiveness in capturing dominant singular values.

\begin{figure}[!h]
  \centering
  \includegraphics[width=8cm]{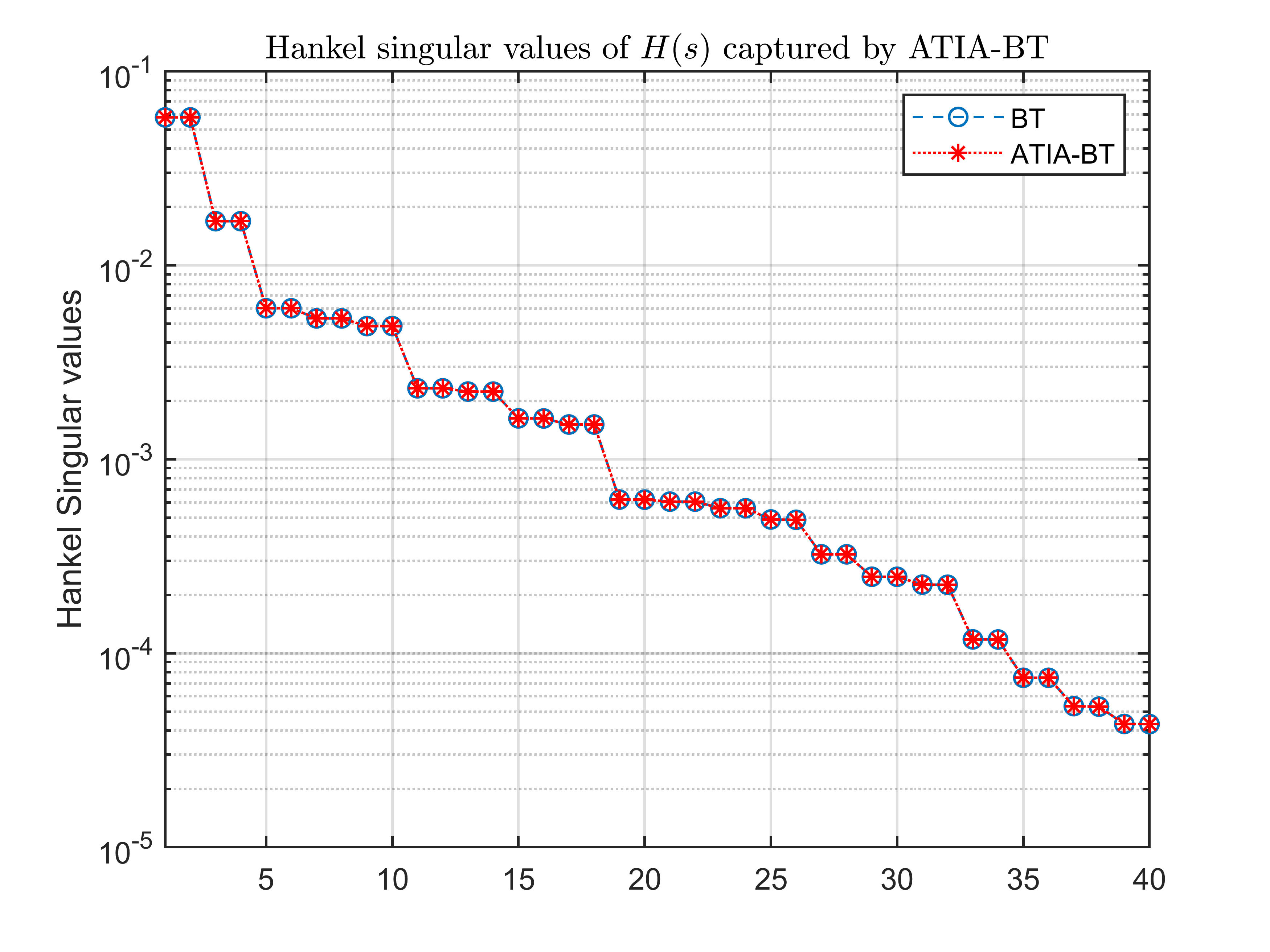}
  \caption{Hankel Singular Values of $40^{th}$-order $H_r(s)$}\label{fig14}
\end{figure}

\begin{figure}[!h]
  \centering
  \includegraphics[width=8cm]{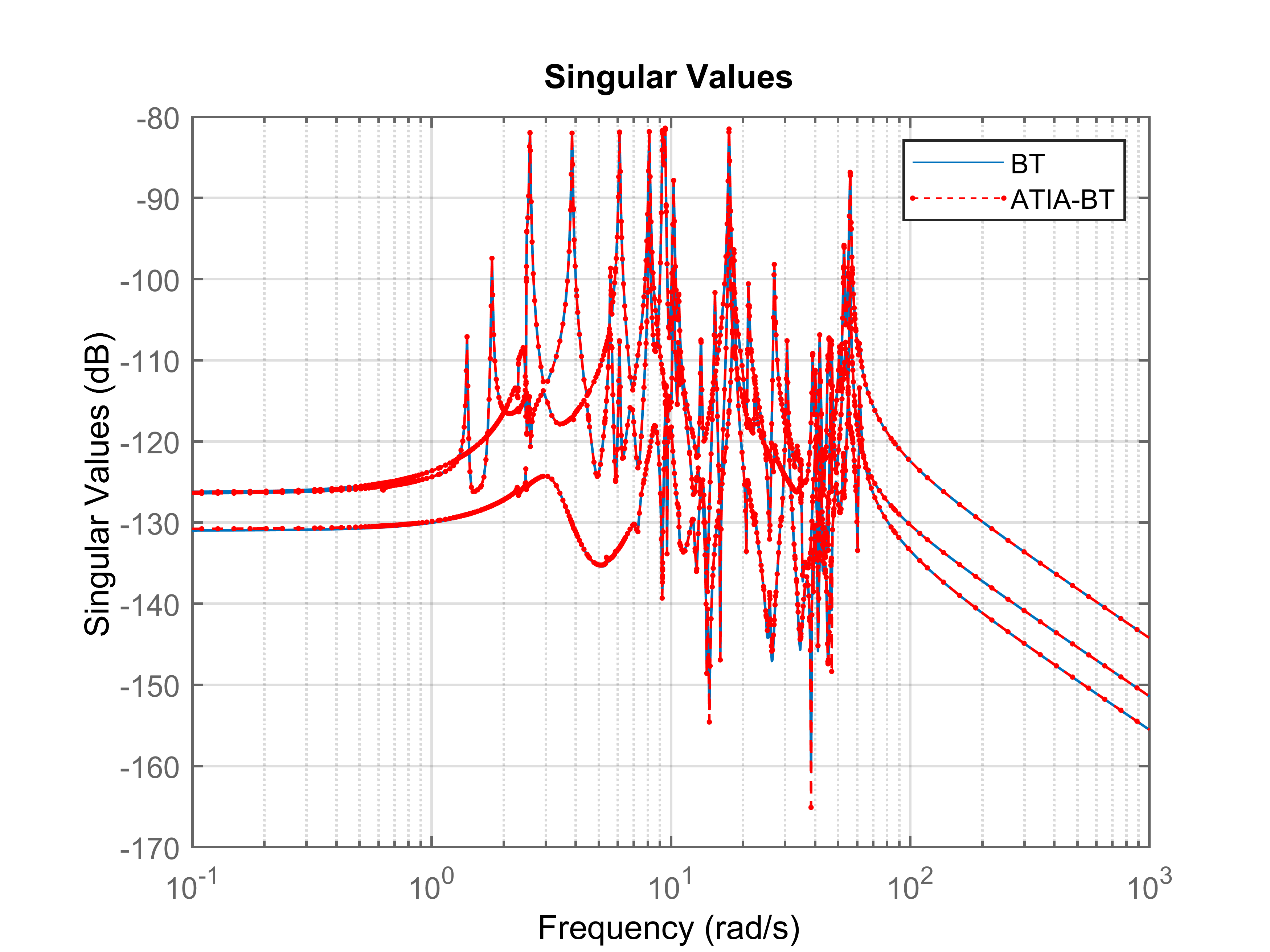}
  \caption{Singular values of $H(s)-H_r(s)$}\label{fig15}
\end{figure}

To evaluate ATIA-BT’s performance, \( tol \) is varied from \( 10^{-3} \) to \( 10^{-5} \), with the relative error \( \frac{\| G(s) - G_r(s) \|_{\mathcal{H}_\infty}}{\| G(s) \|_{\mathcal{H}_\infty}} \) tabulated in Table \ref{tab6}. As \( tol \) decreases, the relative error decreases identical to BT. The Hankel singular values of the $40^{th}$-order ROM generated by ATIA-BT with \( tol = 10^{-3} \) are compared with those from BT in Figure \ref{fig14}, demonstrating ATIA-BT's capability in accurately capturing the $40$ largest Hankel singular values of \( H(s) \). Additionally, Figure \ref{fig15} displays the singular values of \( H(s) - H_r(s) \) for the $40^{th}$-order ROMs obtained by both methods, confirming that ATIA-BT and BT exhibit identical accuracy.
\begin{table}[!h]
\centering
\caption{Comparison of Relative Error $\frac{||G(s)-G_r(s)||_{\mathcal{H}_\infty}}{||G(s)||_{\mathcal{H}_\infty}}$}\label{tab6}
\begin{tabular}{|c|c|c|c|}\hline
$tol$&r&ATIA-BT&BT\\\hline
$10^{-3}$&40&$7.4553\times10^{-4}$&$7.4547\times10^{-4}$\\
$10^{-4}$&50&$3.9230\times 10^{-4}$&$3.9230\times 10^{-4}$\\
$10^{-5}$&50&$3.9230\times 10^{-4}$&$3.9230\times 10^{-4}$\\\hline
\end{tabular}
\end{table}

\textbf{Example 4: Heat Transfer Model}
The heat transfer model is taken from \cite{chahlaoui2005benchmark}, which describes heat transfer through the rod. The partial differential equation provided in \cite{chahlaoui2005benchmark} is discretized with a spatial step size of $\frac{1}{10^7+1}$. The resulting model is a single-input single-output system with an order of ten million, i.e., $10^7$. Since exact Gramian computation is infeasible due to the limited memory resources of the computer used for the experiment, we focus on computational time in this example. Here, the initial order \( r \) is set to 2, with an increment \( \Delta r \) of 2, i.e., \( r = 2 \) and \( \Delta r = 2 \). The maximum allowable iterations are fixed at $i_{max} = 3$ and $k_{max} = 21$ for all experiments on this model. A state-space model of order $2$ is generated using MATLAB’s \textit{rss} command and remains unchanged throughout the experiments. By varying the tolerance $tol$, the elapsed time for computing low-rank approximations of $P$ and $Q$ using ALRS-LYAP is listed in Table 10. The elapsed time for ATIA-BT to produce ROMs is also included in Table 10. Despite the high order of the original model, the algorithms converged for $tol=10^{-4}$ within reasonable time without memory issues.
\begin{table}[!h]
\centering
\caption{Simulation Time Comparison (sec)}\label{tab5}
\begin{tabular}{|c|c|c|c|}\hline
$tol$&ALRS-LYAP $(P)$&ALRS-LYAP $(Q)$&ATIA-BT\\\hline
$10^{-4}$&423.2271 $(r=20)$&497.2445 $(r=20)$&260.8625 $(r=8)$\\
$10^{-5}$&681.9599 $(r=22)$&598.6833 $(r=22)$&1978.3259 $(r=16)$\\\hline
\end{tabular}
\end{table}
\section{Conclusion}
This paper investigates the preservation of the $r$ largest singular values of controllability/observability Gramians and the $r$ largest Hankel singular values of a system through tangential interpolation at $r$ points. The appropriate interpolation points and tangential directions were identified; however, this data is not known \textit{a priori}. To address this, iterative algorithms were developed to automatically determine the necessary interpolation points and tangential directions, ensuring accurate approximation of the dominant singular values of the Gramians and the dominant Hankel singular values of the original system. These algorithms are fully automatic, requiring no user intervention, and adaptively determine the rank of the approximated Gramians as well as the order of the reduced model. Numerical experiments on benchmark problems confirm the theoretical findings, demonstrating the effectiveness of the proposed approach.
\section*{Funding}                               
This work is supported by the National Natural Science Foundation of China under Grants No. 62350410484 and 62273059, and in part by the High-end Foreign Expert Program No. G2023027005L granted by the State Administration of Foreign Experts Affairs (SAFEA).

\end{document}